\documentclass[journal ]{new-aiaa}
\usepackage[utf8]{inputenc}

\usepackage{graphicx}
\usepackage{amsmath}
\usepackage[version=4]{mhchem}
\usepackage{siunitx}
\usepackage{longtable,tabularx}
\usepackage{multirow}
\usepackage{subcaption}
\usepackage{booktabs}
\usepackage{fancyhdr}

\usepackage{nomencl}
  \makenomenclature

\newcommand\norm[1]{\left\lVert#1\right\rVert_2}

\DeclareMathOperator*{\argmin}{arg\,min}
\DeclareMathOperator*{\argmax}{arg\,max}

\newcommand{\iterIdx}{\ensuremath{n}}           
\newcommand{\normVarIdx}{\ensuremath{v}}        
\newcommand{\varIdx}{\ensuremath{i}}            
\newcommand{\genIdx}{\ensuremath{i}}            

\newcommand{\stateVar}{\ensuremath{q}}
\newcommand{\solConsFOM}{\ensuremath{\mathbf{\stateVar}}}
\newcommand{\solPrimFOM}{\ensuremath{\mathbf{\stateVar}_p}}
\newcommand{\solPrimFOMAdaptive}{\ensuremath{\hat{\mathbf{\stateVar}}_p}}
\newcommand{\solConsFOMFunc}[1]{\ensuremath{\solConsFOM \left( #1 \right)}}

\newcommand{\solConsFOMRef}{\ensuremath{\mathbf{\stateVar}_{\text{ref}}}}
\newcommand{\solPrimFOMRef}{\ensuremath{\mathbf{\stateVar}_{p,\text{ref}}}}

\newcommand{\solPrimROMFull}{\ensuremath{\widetilde{\mathbf{\stateVar}}_p}}

\newcommand{\solPrimROMRed}{\ensuremath{\mathbf{\stateVar}}_{r}}
\newcommand{\solPrimFOMVar}{\ensuremath{\mathbf{\stateVar}_{p,\varIdx}}}

\newcommand{\solPrimROMProj}{\ensuremath{\mathbf{\bar{\stateVar}}_{p}}}
\newcommand{\solPrimROMProjVar}{\ensuremath{\mathbf{\bar{\stateVar}}_{p,\varIdx}}}

\newcommand{\rhsVar}{\ensuremath{f}}
\newcommand{\rhs}{\ensuremath{\mathbf{\rhsVar}}}
\newcommand{\rhsFunc}[1]{\ensuremath{\rhs \left( #1 \right)}}

\newcommand{\resVar}{\ensuremath{r}}
\newcommand{\res}{\ensuremath{\mathbf{\resVar}}}
\newcommand{\resFunc}[1]{\ensuremath{\res \left( #1 \right)}}

\newcommand{\scaleVarCons}{\ensuremath{P}}
\newcommand{\scaleMatCons}{\ensuremath{\mathbf{\MakeUppercase{\scaleVarCons}}}}
\newcommand{\scaleVarPrim}{\ensuremath{H}}
\newcommand{\scaleMatPrim}{\ensuremath{\mathbf{\MakeUppercase{\scaleVarPrim}}}}

\newcommand{\trialBasisVar}{\ensuremath{V}}

\newcommand{\trialBasisPrim}{{\ensuremath{\mathbf{\MakeUppercase{\trialBasisVar}}_p}}}

\newcommand{\trialBasisPrimN}[1]{{\ensuremath{\mathbf{\MakeUppercase{\trialBasisVar}}^{#1}_p}}}
\newcommand{\testBasisVar}{\ensuremath{W}}

\newcommand{\testBasisPrim}{\ensuremath{{\mathbf{\MakeUppercase{\testBasisVar}}}_{{p}}}}

\newcommand{\testBasisPrimGPOD}{\ensuremath{{\mathbf{\overline{\MakeUppercase{\testBasisVar}}}}_{{p}}}}

\newcommand{\resApprox}{\ensuremath{\mathbf{\overline{\resVar}}}}

\newcommand{\sampVar}{\ensuremath{S}}
\newcommand{\sampMat}{\ensuremath{\mathbf{\MakeUppercase{\sampVar}}}}

\newcommand{\numDOF}{\ensuremath{N}}            
\newcommand{\numElements}{\ensuremath{N_{elem}}}    
\newcommand{\numVars}{\ensuremath{N_{var}}}     
\newcommand{\numSolModes}{\ensuremath{n_p}}     
\newcommand{\numSolModesTotal}{\ensuremath{n_{p,total}}}
\newcommand{\numSamps}{\ensuremath{n_s}}

\newcommand{\timeVar}{\text{t}}                 
\newcommand{\dt}{\ensuremath{\Delta \text{t}}}  
\newcommand{\dTimeVar}{\text{dt}}               

\newcommand{\errProjVsFOM}{\ensuremath{\bar{\epsilon}}}

\newcommand{\lp}{\left(}
\newcommand{\rp}{\right)}

\newcommand{\zeroVec}{\ensuremath{\mathbf{0}}}

\newcommand{\addrTwo}[1]{\textcolor{black}{#1}}

\title{Investigations on Projection-Based Reduced Order Model Development for Rotating Detonation Engine\footnote{Presented at AIAA SciTech Forum in Orlando, FL on January 11, 2024: AIAA-2024-2038}}

\author{Ryan Camacho \footnote{Doctoral Student, Aerospace Engineering, G410 LEEP2, 1536 W 15th St, Lawrence, KS 66045} and Cheng Huang\footnote{Assistant Professor, Aerospace Engineering, 2118D Learned Hall, 1530 W 15th St, Lawrence, KS 66045}}

\affil{University of Kansas, Lawrence, KS, 66045}

\begin{document}

\maketitle

\begin{abstract}
The current study aims to evaluate and investigate the development of projection-based reduced-order models (ROMs) for efficient and accurate RDE simulations. Specifically, we focus on assessing the projection-based ROM construction utilizing three different approaches: the linear static basis, nonlinear quadratic basis, and an adaptive model order reduction (MOR) formulation. First, an ~\textit{a priori} analysis is performed to evaluate the effectiveness of the linear static and nonlinear quadratic bases in representing the detonation-wave dynamics. The~\textit{a priori} analysis reveals that compared to the linear basis, the nonlinear quadratic basis provides significantly improved representation of detonation-wave dynamics within the training regime. However, it exhibits limited capabilities in representing the dynamics beyond the training regime, either in the future state or under a different operating parameter (i.e., inlet velocity). Second, the investigations proceed to the adaptive MOR formulation, which constructs an \textit{online} adaptive ROM with a small amount of offline training data. It is  demonstrated that the adaptive ROM can provide significantly enhanced predictive capabilities in modeling the RDE dynamics in the future state, and subject to parametric variations. More importantly, the adaptive ROM is shown to be capable of capturing the initial transience in establishing the detonation wave.
\end{abstract}
\section{Introduction}
\label{sec:intro}
\lettrine{P}ressure gain combustion is recognized as a viable alternative to traditional combustion in propulsion applications due to its theoretical performance improvements~\cite{Adamson_1967,Zeldovich_1940}. As one of the most prominent realizations based on this concept, the rotating detonation engine (RDE) offers several advantages for application in propulsion or land-based power generation, including mechanical simplification, broad operability limits~\cite{Anand_2016,Fotia_2016}, potential to increase thermal efficiency~\cite{Nordeen_2014,Shao_2010}, and the reduction of propellant pumping requirements~\cite{Sousa_2017,Rankin_2017}. Therefore, RDEs have garnered significant interest for both experimental~\cite{Bennewitz_2019,Kindracki_2011,Walters_2021} and numerical~\cite{Pal_LESRDE,Prakash_2021,Lietz_2020,Schwer_2019,Sato_2021_MassFlow} investigations to better understand the underlying physics. However, the practical applications of RDEs require a considerable number of parametric evaluations over a range of operating conditions and geometric designs. Conducting such a large number of experiments consumes significant resources and provides restricted access to quantify and understand the underlying physics. On the other hand, high-fidelity simulations, such as large-eddy simulations (LES), can provide valuable physical insights for RDEs. Even with the modern exascale computing capabilities~\cite{Rood_et_al}, however, due to the inherent physics complexity, these simulations remain prohibitively expensive for practical engineering applications. For example, Lietz et al.~\citep{Lietz_2019} illustrated that a three-dimensional (3D) LES of an RDE was feasible, but even with a subset of the experimental geometry, approximately 0.18 million CPU hours were required to simulate 2ms of engine run-time. While engineering designs often require hundreds of evaluations through these simulations, they are not affordable for practical RDE applications.

To address this gap, reduced models, either physics-based~\cite{Kaemming_2017} or data-driven~\cite{Koch_2021}, have been developed in the literature to provide efficient and reasonably accurate predictions of performance metrics. Physics-based reduced models typically seek to reduce the computational cost through simplification of the physical models, and are referred to as reduced-fidelity models (RFMs). A common approach for RDE applications is to model a 3D configuration as an unrolled two-dimensional (2D) domain with periodic boundary conditions, which has been widely applied and demonstrated effective for premixed RDE applications ~\citep{Zhao_2021,Schwer_2019}. Additional reductions in modeling fidelity can further increase computational efficiency. Examples include 1) solving 2D Euler instead of 2D Navier-Stokes equations ~\citep{Mahmoudi_2014}; 2) solving 1D Euler equations with a modeled source term to approximate second-dimension injection, mixing, and exhaustion as a lump volume~\cite{Koch_2020ModeLocked}; 3) developing a two-equation phenomenological modeling framework  for the detonation wave dynamics based on experimental observations~\citep{Koch_2021RDE}; or 4) constructing a simple thermodynamic model to predict the quantities of interest~\cite{Kaemming_2017}.

Data-driven reduced models, on the other hand, seek to inform reduced-order models (ROMs) using either simulation and experiment data instead of relying on physics simplifications. Such methods can be classified as either machine learning (ML) or model-order reduction (MOR). ML is a pure data-driven black-box approach that estimates associations between system inputs, outputs, and parameters using a limited number of observations ~\citep{cherkassky2007learning}, leading to a low entry barrier for implementations. For example, Koch~\cite{Koch_2021} trains neural ordinary differential equations from high-speed camera footage to approximate the Arrhenius kinetics and the mixing models in the two-equation phenomenological modeling framework~\cite{Koch_2020}. Mendible et al.~\cite{Mendible_2021} demonstrated the feasibility of constructing ROMs for RDE directly from experimental data. However, one challenge is that ML methods lacks generalizability ~\citep{Zhou_2022,Duraisamy_2021,Ihme_2022} - i.e., the trained ML models often need to be specially trained and are only applicable to a narrow range of applications. As a result, broadly applicable ML models would require copious amounts of data for training, resulting in large data storage requirements. Additionally, ML models will not always satisfy physical laws (e.g., conservation of mass) as this method only seeks to create a black box that relates the inputs and outputs. While there has been recent research into incorporating physical laws in ML \citep{Raissi_2019}, its applicability to physically complex problems such as turbulent reacting flows remains an open question. 

The second data-driven approach, MOR (specifically, projection-based MOR) derives ROMs to represent the evolution of a high-dimensional nonlinear PDE by projecting the PDE onto a low-dimensional subspace ~\citep{Benner_2015}. the core factor that distinguishes MOR from ML is the direct integration of PDE (i.e., physics) into the reduced model, ensuring that the MOR-ROM respects the model physics and produces physically accurate results. The importance of the integration of PDE has been extensively documented ~\citep{Peherstorfer_2016,Kramer_2019,McQuarrieOpInf2021}. MOR has been used successfully to build ROMs for a broad range of applications ~\citep{Barbagallo_2012,Lieu_2007,Blonigan_2021,HuangMPLSVT2022} and, more recently, Farcas et al.~\citep{Farcas_2022,farcas2023improving} successfully used a non-intrusive MOR method to model a coarsened RDE problem. Typically, the construction of a projection based ROM consists of two stages: 
\begin{itemize}
    \item
     An \textit{offline} stage that performs full-order-model (FOM) simulations of the target systems to construct low-dimensional subspaces to represent the high dimensional solution data, typically through proper orthogonal decomposition (POD)~\cite{Lumley1997}. 

    \item
    And \textit{online} stage that executes ROM through evolving the dynamics on the low-dimensional spaces by projecting the FOM onto the lower dimensional subspace. 
\end{itemize}

However, projection-based ROM development for RDE applications present one major challenge, Detonation-wave dynamics feature convection-dominated multi-scale physics and exhibit a slow decay of Kolmogorov N-width~\citep{Greif_2019} that cannot be effectively approximated using low-dimensional subspaces (e.g., those by POD). Thus, these problems are typically not amenable to data reduction. This specific issue is referred to as the Kolmogorov barrier. As a result, when greater accuracy is desired, it becomes necessary to greatly increase the dimensionality of the low dimensional subspace, increasing the computational expense of the resulting ROM.

In order to break the Kolmogorov barrier, different schools of thoughts have arisen. The first strategy seeks to construct multiple local subspaces to describe the high-dimensional convection-dominated dynamics, as opposed to a single global subspace ~\citep{Geelen_2022,Amsallem_2015}. In this method the ROM can be adapted to local dynamics rather than trying to resolve the entire domain, allowing for decreased dimensionality and better performance. The second strategy uses nonlinear bases (or manifolds) to construct the low-dimensional subspaces. One class of popular approaches leverages subspace/snapshot transformation to construct such basis, with the goal of recovering low-rank structures with rapid-decay Kolmogorov N-widths. These approaches include the method of freezing ~\citep{Ohlberger_2013}, shifted POD ~\citep{Reiss_2018,schulze2018model,Reiss_2021}, transported subspaces ~\citep{Rim_2023} or snapshots ~\citep{Nair_2019} , and implicit-feature tracking~\citep{Mirhoseini_2023}. Other researchers directly compute nonlinear basis using convolutional autoencoders~\citep{KimChoiNonlinearMainfold,LeeCarlbergNonlinearManifold} or quadratic approximation~\citep{Geelen_2023_Quadratic,Barnett_QuadraticPROM2022}. The third strategy is to use adaptive MOR ~\citep{Peherstorfer_2022_Kolmogorov,Zucatti_2024} to break the Kolmogorov barrier by updating the ROM in the online stage to either satisfy a posteriori error estimation or seek an optimal representation of the target problems’ evaluated dynamics. For example, Peherstorfer~\citep{PeherstorferADEIM} adapts the affine approximation space by exploiting the locality in space and time of propagating coherent structures evaluated from the sampled FOM solutions. Ramezanian et al.~\citep{Ramezanian_2021} proposed an on-the-fly ROM method for reactive-species transport by deriving evolution equations for a low-dimensional time-dependent manifold through a variational principle. 

Therefore, the objective of the current paper is to investigate existing methods of projection-based ROM development for RDE applications. Specifically, we consider three approaches to construct the projection-based ROM: 1) the linear static basis, 2) the nonlinear quadratic basis~\citep{Geelen_2023_Quadratic}, and 3) adaptive MOR~\citep{Huang_2023}. Detailed assessments of the three ROM methods are performed using a benchmark 2D RDE problem, which exhibits distinct detonation wave dynamics, featuring slow decay of Kolmogorov N-width, subject to variations in operating conditions. The remainder of the paper is organized as follows. Section~\ref{sec:fom} presents the FOM formulation. Section~\ref{sec:rom} presents the ROM formulation, including the construction of the linear and nonlinear quadratic bases, and a recently developed adaptive ROM formulation. Section~\ref{sec:configuration} discusses the computational configuration for the RDE problem. Section~\ref{sec:results} presents the numerical results and investigations based on the 2D RDE configuration including the \textit{a priori} analysis for assessment of the linear and nonlinear quadratic bases and the evaluations of the adaptive ROM. In Section~\ref{sec:conclusion}, we provide concluding remarks and future directions for the work.
\section{Full-Order Model}
\label{sec:fom}

The governing equations of the full-order model (FOM) used for all the numerical results in the current paper can be expressed in a generic semi-discrete form
\begin{equation}
    \frac{\text{d} \solConsFOM \left(\solPrimFOM \right)}{\dTimeVar} = \rhsFunc{\solPrimFOM} \ \ \ \text{with} \ \ \ \solPrimFOM \left( \timeVar = 0 \right) = \solPrimFOM^0,
     \label{eq:fom_semi_discrete}
\end{equation}
where $\timeVar \in \left[0,t_f \right]$ is the solution time, which spans the time interval from $0$ to $t_f$, $\solPrimFOM \in {\mathbb{R}^{\numDOF}}$ is the vector of solution (or state) variables, $\solPrimFOM^0 \in {\mathbb{R}^{\numDOF}}$ is the vector of states to be specified as the initial conditions at $t = 0$, $\solConsFOM: \mathbb{R}^{\numDOF} \rightarrow \mathbb{R}^{\numDOF}$ and $\rhs: \mathbb{R}^{\numDOF} \rightarrow \mathbb{R}^{\numDOF}$ are nonlinear functions of $\solPrimFOM$. $\numDOF$ is the total number of degrees of
freedom in the system. The function $\solConsFOM$ represents the conservative state in the case of a FOM based on conservation laws, and the function $\rhs$ represents surface fluxes, source terms, and body forces resulting from the spatial discretization of the governing equations. In addition, different time-discretization methods can be introduced to solve Eq.~\ref{eq:fom_semi_discrete}, such as linear multi-step, or Runge--Kutta methods~\citep{ButcherNumMeth}. For the analysis performed in this paper, a second order implicit scheme was used for both FOM and ROM calculations
\begin{equation}
    \resFunc{\solPrimFOM^\iterIdx} \triangleq \frac{3}{2} \solConsFOMFunc{\solPrimFOM^{\iterIdx}} - 2 \solConsFOMFunc{\solPrimFOM^{\iterIdx-1}} + \frac{1}{2} \solConsFOMFunc{\solPrimFOM^{\iterIdx-2}} - \dt \rhsFunc{\solPrimFOM^\iterIdx} = 0.
    \label{eq:FOMresidual}
\end{equation}
where $\dt \in \mathbb{R}^{+}$ is the physical time step for the numerical solution, and $\res: \mathbb{R}^{\numDOF} \rightarrow \mathbb{R}^{\numDOF}$ is defined as the FOM equation residual. The state variables, $\solPrimFOM^\iterIdx$, are solved for at each time step so that $\resFunc{\solPrimFOM^\iterIdx} = \zeroVec$. the details of the discretization can be found in Ref.~\citep{HuangMPLSVT2022}.

\section{Reduced-Order Model}
\label{sec:rom}
As discussed in Section~\ref{sec:intro}, the ROM construction for problems featuring multi-scale phenomena with strong convection and nonlinear effects is well-recognized to be a major challenge due to the slow decay of Kolmogorov N-width. Therefore, we investigate two methods to address this challenge: 1. using nonlinear quadratic basis to construct low-dimensional subspace~\citep{Geelen_2023_Quadratic,Barnett_QuadraticPROM2022}, and 2. an adaptive ROM formulation developed by Huang and Duraisamy~\citep{Huang_2023}. In this section, we first briefly describe the low-dimensional subspace construction via linear basis and then introduce the two methods.

\subsection{Construction of Low-dimensional Subspace for Solution Variables using Linear Basis}
\label{subsec:pod}
The state variables, $\solPrimFOM$ in Eq.~\ref{eq:fom_semi_discrete}, are approximated using a set of low-dimensional subspace (or trial basis), $\trialBasisPrim \in \mathbb{R}^{\numDOF \times \numSolModes}$
\begin{equation}
    \scaleMatPrim \lp \solPrimROMFull - \solPrimFOMRef \rp = \trialBasisPrim \solPrimROMRed.
    \label{eq:pod_qp}
\end{equation}
where $\solPrimROMFull$ represents an approximation of the state variable (i.e. $\solPrimROMFull \approx \solPrimFOM$). $\solConsFOMRef$ represents a reference state, which is selected to be the initial FOM solution $\solPrimFOMRef = \solPrimFOM(\timeVar = \timeVar_0)$ in the training data for the current investigations. The trial basis$\trialBasisPrim$ is computed via POD ~\citep{lumley1967structure,Berkooz_1993} using the singular value decomposition (SVD). $\solPrimROMRed \in \mathbb{R}^{\numSolModes}$ is the reduced state with $\numSolModes$ representing the number of trial basis. $\scaleMatPrim \in{\mathbb{R}^{\numDOF \times \numDOF}}$ represents a scaling matrix, introduced to keep all variables of similar orders of magnitude when computing the trial basis. In the current study, all quantities are normalized  by their $L^2$-norm, as proposed by Lumley and Poje~\citep{Lumley1997}
\begin{equation}
    \scaleMatPrim = diag \lp \scaleMatPrim_1, \ldots, \scaleMatPrim_i, \ldots, \scaleMatPrim_{\numElements} \rp,
    \label{eq:pod_norm}
\end{equation} 
where $\scaleMatPrim_i = diag\left( \phi^{-1}_{1,norm}, \ldots , \phi^{-1}_{\numVars,norm} \right)$. Here, $\phi_{\normVarIdx,norm}$ represents the $\normVarIdx^{th}$ state  variable \addrTwo{in $\solPrimFOM$} and
\begin{equation}
    \phi_{\normVarIdx,norm} = {\frac{1}{\timeVar_1 - \timeVar_0}\int^{\timeVar_1}_{\timeVar_0} \frac{1}{\Omega} \int_{\Omega} \phi'^2_\normVarIdx(\mathbf{x}, \timeVar) \; \text{d}\mathbf{x} \; \text{dt}}.
    \label{eq:l2norm}
\end{equation}

\subsection{Construction of Low-dimensional Subspace for Solution Variables using Nonlinear Quadratic Basis}
Alternatively, nonlinear basis is pursued to construct the low-dimensional subspace for improved approximation of the state variables, $\solPrimFOM$. Specifically, in the current paper, we seek to construct such nonlinear subspace by introducing an additional quadratic structure to Eq.~\ref{eq:pod_qp} following the work by Geelen et al.~\citep{Geelen_2023_Quadratic}

\begin{equation}
    \scaleMatPrim \lp \solPrimROMFull - \solPrimFOMRef \rp = \trialBasisPrim \solPrimROMRed + \Bar{\mathbf{V}}_{p}(\solPrimROMRed \otimes \solPrimROMRed).
    \label{eq:pod_nonlinear}
\end{equation}
where $\Bar{\mathbf{V}}_{p} \in {\mathbb{R}^{\numDOF \times n^2_p}}$ represents a quadratic mapping operator and $\otimes$ denotes the Kronecker product. The quadratic operator $\Bar{\mathbf{V}}_{p}$ is computed from the following optimization problem
\begin{equation}
    \Bar{\mathbf{V}}_{p} \triangleq \argmin_{\Bar{\mathbf{V}}_{p} \in {\mathbb{R}^{\numDOF \times n^2_p}}} \norm{ \scaleMatPrim \lp \solPrimFOM - \solPrimFOMRef \rp - \trialBasisPrim \solPrimROMRed + \Bar{\mathbf{V}}_{p}(\solPrimROMRed \otimes \solPrimROMRed) }^2.
    \label{eq:nonlinear_vbar}
\end{equation}
In addition, as reported by Barnett and Farhat~\citep{Barnett_QuadraticPROM2022} and Geelen et al.~\citep{Geelen_2023_Quadratic}, directly solving $\Bar{\mathbf{V}}_{p}$ from Eq.~\ref{eq:nonlinear_vbar} is prone to overfitting the noise in the training data and it is necessary to introduce regularization to the least-square minimization problem above. Therefore, we follow the approach by Geelen et al.~\citep{Geelen_2023_Quadratic} to introduce a regularization term to Eq.~\ref{eq:nonlinear_vbar}
\begin{equation}
    \Bar{\mathbf{V}}_{p} \triangleq \argmin_{\Bar{\mathbf{V}}_{p} \in {\mathbb{R}^{\numDOF \times n^2_p}}} \norm{ \scaleMatPrim \lp \solPrimFOM - \solPrimFOMRef \rp - \trialBasisPrim \solPrimROMRed + \Bar{\mathbf{V}}_{p}(\solPrimROMRed \otimes \solPrimROMRed) }^2 + \lambda \norm{\Bar{\mathbf{V}}_{p}}^2.
    \label{eq:nonlinear_vbar_reg}
\end{equation}
where $\lambda$ is a scalar regularization parameter. We refer the readers to the work by Geelen et al.~\citep{Geelen_2023_Quadratic} for details on computing $\Bar{\mathbf{V}}_{p}$. It is remarked that this quadratic nonlinear subspace has been proposed and demonstrated to provide significant improvements on ROM performance over the linear subspace in model order reduction for both non-intrusive method by Geelen et al.~\citep{Geelen_2023_Quadratic} and intrusive method by Barnett and Farhat~\citep{Barnett_QuadraticPROM2022}. On the other hand, we also remark the dimension of the operator $\Bar{\mathbf{V}}_{p}$ scales quadratically with the number of trial basis, which incurs significant additional cost in offline and online computations compared to its counterpart linear basis. In the current study, we focus on evaluating the characteristics of the quadratic nonlinear basis in representing the detonation wave dynamics in RDEs based on an offline \textit{a prior} assessment, which is discussed in Section \ref{sec:apriori}.

\subsection{Adaptive ROM Formulation}
\label{sec:adaptivity}
Different from the conventional offline-online ROM methods with static trial basis, the adaptve ROM method aims to update both the trial basis, $\trialBasisPrim$, during the online ROM calculation. Following the work by Huang and Duraisamy~\citep{Huang_2023}, the following minimization problem is posed to construct the adaptive ROM
\begin{equation}
    \{ \solPrimROMRed^n, \trialBasisPrimN{n}, \sampMat^{n} \} \triangleq \argmin_{ \solPrimROMRed \in \mathbb{R}^{\numSolModes} , \trialBasisPrimN{\iterIdx} \in \mathbb{R}^{\numDOF \times \numSolModes} , \sampMat^{T}_{n} \in \mathbb{S}^{\numDOF \times \numSamps } } \norm{\trialBasisPrimN{\iterIdx} \left[ \sampMat^{T}_{n} \trialBasisPrimN{\iterIdx} \right]^{+} \sampMat^{T}_{n} \scaleMatCons \resFunc{\solPrimROMFull^{\iterIdx}}}^2 
    \label{eq:adaptiverom:def}
\end{equation}
where $\solPrimROMFull^{\iterIdx} = \solPrimFOMRef + \scaleMatPrim^{-1} \trialBasisPrimN{\iterIdx} \solPrimROMRed^{\iterIdx}$ and $\res$ is the nonlinear FOM equation residual defined in Eq.~\ref{eq:FOMresidual}. To circumvent the computational bottleneck in evaluating the nonlinear function $\res$, hyper-reduction~\citep{Everson1995_gappyPOD} is applied so that a full reconstruction of $\res$ can be obtained using a small number of $\numSamps$ sparse samples (i.e., $\numSamps << \numDOF$), $\resApprox^n \approx \trialBasisPrimN{\iterIdx} \lp \sampMat^{T}_{n} \trialBasisPrimN{\iterIdx} \rp^+  \sampMat^{T}_{n} \res^n$. $\sampMat^T \in \mathbb{S}^{\numSamps \times \numDOF}$ represents a selection operator that belongs to a class of matrices with $\numSamps$ columns (i.e., sampling points) of
the identity matrix, $\mathbf{I} \in \mathbb{I}^{\numDOF \times \numDOF}$ and $\sampMat$ is also constantly updated in adaptive ROM. However, directly solving Eq.~\ref{eq:adaptiverom:def} is computationally intractable and therefore a decoupled approach based on a predictor-corrector idea is utilized instead in the following three steps. 

\textbf{First}, $\solPrimROMRed^{\iterIdx}$ is selected as the solution to the minimization problem below following  a model-form preserving least-squares with variable transformation (MP-LSVT) formulation developed by Huang et al.~\citep{HuangMPLSVT2022}
\begin{equation}
    \solPrimROMFull^{\iterIdx} \triangleq \argmin_{\solPrimROMFull^{\iterIdx} \in \textrm{Range}(\trialBasisPrim)} \norm{\trialBasisPrimN{\iterIdx-1} \lp \sampMat^{T}_{n-1} \trialBasisPrimN{\iterIdx-1} \rp^{+} \sampMat^{T}_{n-1} \scaleMatCons \resFunc{\solPrimROMFull^{\iterIdx}}}^2,
    \label{eq:mplsvt_hyper}
\end{equation}
with $\solPrimROMFull^{\iterIdx} = \solPrimFOMRef + \scaleMatPrim^{-1} \trialBasisPrimN{\iterIdx-1} \solPrimROMRed^{\iterIdx}$, leading to a reduced non-linear system of dimension $\numSolModes$ which can be viewed as the result of a Petrov-Galerkin projection
\begin{equation}
    (\testBasisPrimGPOD^\iterIdx)^T \trialBasisPrimN{\iterIdx-1} \lp \sampMat^{T}_{n-1} \trialBasisPrimN{\iterIdx-1} \rp^{+} \sampMat^{T}_{n-1} \scaleMatCons \resFunc{\solPrimROMFull^{\iterIdx}} = \zeroVec.
    \label{eq:mplsvt_proj_hyperS}
\end{equation}
where $\testBasisPrimGPOD^\iterIdx$ is the resulting test basis
\begin{equation}
     \testBasisPrimGPOD^\iterIdx = \trialBasisPrimN{\iterIdx-1} \lp \sampMat^T_{n-1}  \trialBasisPrimN{\iterIdx-1} \rp^{+} \sampMat^T_{n-1} \testBasisPrim^\iterIdx.
     \label{eq:mplsvt_w_hyper}
\end{equation}

\textbf{Second}, the trial basis, $\trialBasisPrim$, is updated by solving the minimization problem 
\begin{equation}
    \trialBasisPrimN{n}  \triangleq  \argmin_{ \trialBasisPrimN{\iterIdx} \in \mathbb{R}^{\numDOF \times \numSolModes} } \norm{ \scaleMatCons \resFunc{\solPrimROMFull^{\iterIdx}}}^2.
    \label{eq:adaptiverom:basis_minimization}
\end{equation}
As proved by Huang and Duraisamy~\citep{Huang_2023}, Eq.~\ref{eq:adaptiverom:basis_minimization} can be solved exactly via the update
\begin{equation}
    \trialBasisPrimN{n} = \trialBasisPrimN{n-1} + \delta \trialBasisPrim,
    \label{eq:adaptiverom:basis_update}
\end{equation}
through an increment, $\delta\trialBasisPrim \in \mathbb{R}^{\numDOF \times \numSolModes}$, given by
\begin{equation}
    \delta \trialBasisPrim = \frac{ \lp \solPrimFOMAdaptive^n - \solPrimROMFull^n \rp (\solPrimROMRed^n)^T}{||\solPrimROMRed^n||_2^2},
    \label{eq:adaptiverom:delta}
\end{equation}
where $\solPrimFOMAdaptive^{\iterIdx} \in \mathbb{R}^{\numDOF}$ represents the full-state information evaluated based on the FOM equation residual as follows
\begin{equation}
    \resFunc{\solPrimFOMAdaptive^\iterIdx} = \frac{3}{2} \solConsFOMFunc{\solPrimFOMAdaptive^{\iterIdx}} - 2 \solConsFOMFunc{\solPrimROMFull^{\iterIdx-1}} + \frac{1}{2} \solConsFOMFunc{\solPrimROMFull^{\iterIdx-2}} - \dt \rhsFunc{\solPrimFOMAdaptive^\iterIdx, \timeVar^\iterIdx}  = 0,
    \label{eq:adaptiverom:surrogate_fom}
\end{equation}
Here, an alternate formulation to the one in Ref.~\citep{PeherstorferADEIM} is adopted by updating the basis based on the full-state information, $\solPrimFOMAdaptive^{\iterIdx}$, evaluated at the current time step, $n$ instead of collecting at multiple time steps, which is similar to the work done by Zimmermann et al.~\citep{ZimmermannAdaptiveBasis2018}. This formulation is referred to as the one-step adaptive-basis approach. It is worth pointing out that the time step of basis adaptation, denoted as $z_b$, is empirically determined for the target applications and in the current work, the basis is chosen to be adapted at each time step (i.e., $z_b = 1$). 

\textbf{Third}, the sampling points, $\sampMat$, are adapted after the basis adaptation (Eq.~\ref{eq:adaptiverom:basis_update}) to maximize the target quantity of interest $\mathbf{\Theta}$
\begin{equation}
    \sampMat^n  \triangleq  \argmax_{ \sampMat \in \mathbb{S}^{\numDOF \times \numSamps } } \norm{ \mathbf{\Theta} }^2,
    \label{eq:adaptiverom:sampling_minimization}
\end{equation}
where $\mathbf{\Theta} = \norm{ \nabla p } $ is selected to be the magnitude of the pressure gradients. In practice, the sampling points are updated and selected at the first $n_s$ points corresponding to the highest magnitudes of $\mathbf{\Theta}$. It should be noted that the sampling points adaptation (Eq.~\ref{eq:adaptiverom:sampling_minimization}) requires the evaluation of full-state information, $\solPrimFOMAdaptive^\iterIdx$, at all the points, leading to a computationally expensive step. Therefore, to achieve significant gains in computational efficiency for the adaptive ROM formulation, the time step of updating the sampling points denoted as $z_s$ is required to be set as larger value and when the sampling points are not being updated, the basis is updated only at the sampled points
\begin{equation}
    \sampMat^T_{n-1} \trialBasisPrimN{n} = \sampMat^T_{n-1} \trialBasisPrimN{n-1} + \sampMat^T_{n-1} \delta \trialBasisPrim,
    \label{eq:adaptiverom:basis_update_s}
\end{equation}
which only requires the full-state information to be evaluated at the sampled points, significantly reducing the computational cost
\begin{equation}
    \sampMat^T_{n-1} \resFunc{\solPrimFOMAdaptive^\iterIdx} = \sampMat^T_{n-1} \left[ \frac{3}{2} \solConsFOMFunc{\solPrimFOMAdaptive^{\iterIdx}} - 2 \solConsFOMFunc{\solPrimROMFull^{\iterIdx-1}} + \frac{1}{2} \solConsFOMFunc{\solPrimROMFull^{\iterIdx-2}} - \dt \rhsFunc{\solPrimFOMAdaptive^\iterIdx, \timeVar^\iterIdx} \right] = 0,
    \label{eq:adaptiverom:surrogate_fom_s}
\end{equation}

If the sampling points are being updated, the basis is updated at all the points following Eq.~\ref{eq:adaptiverom:basis_update} with the full-state information evaluated at the sampled points, $\sampMat_{n-1}$, using Eq.~\ref{eq:adaptiverom:surrogate_fom_s} while the full-state information evaluated at the unsampled points, $\sampMat^{*}_{n-1}$, are evaluated using a larger time step (i.e. $z_s \dt$) to incorporate the non-local in basis adaptation, which has been concluded as a crucial step in the adaptive ROM development by Huang and Duraisamy~\citep{Huang_2023}.
\begin{equation}
    \sampMat^{*T}_{n-1} \resFunc{\solPrimFOMAdaptive^\iterIdx} = \sampMat^{*T}_{n-1} \left[ \frac{3}{2} \solConsFOMFunc{\solPrimFOMAdaptive^{\iterIdx}} - 2 \solConsFOMFunc{\solPrimROMFull^{\iterIdx-z_s}} + \frac{1}{2} \solConsFOMFunc{\solPrimROMFull^{\iterIdx-2z_s}} - z_s \dt \rhsFunc{\solPrimFOMAdaptive^\iterIdx, \timeVar^\iterIdx} \right] = 0.
    \label{eq:adaptiverom:surrogate_fom_us}
\end{equation}
Similar to $z_b$, the determination of $z_s$ is empirical and can impact the prediction accuracy of the resulting adaptive ROMs as well. It needs to be pointed out that compared to the conventional projection-based ROM methods that leverage \emph{linear static} basis, the adaptive ROM formulation requires additional computational operations (e.g., Eq.~\ref{eq:adaptiverom:surrogate_fom}) to update the basis and sampling points, possibly resulting in fewer gains in computational efficiency. However, the adaptive ROM formulation also presents several major advantages over other methods: 1) it imposes much lower requirements in \emph{offline} training, significantly reducing the computational costs to construct ROMs for complex fluid problems, the high-fidelity simulations of which are often expensive; 2) it usually requires small number of basis modes and inherently addresses the restrictions of slow \emph{Kolmogorov N-width} decay on the \emph{linear static} basis; and 3) more importantly, it significantly improves the predictive capabilities of the ROM as demonstrated by the current author~\citep{Huang_CBROM2022,Huang_2023}, especially in enabling accurate predictions of parametric variations and transient dynamics, which are challenging for other ROM methods.

\section{Computational Configuration and Model}
\label{sec:configuration}
\begin{figure}[hbt!]
\centering
\includegraphics[width=.8\textwidth]{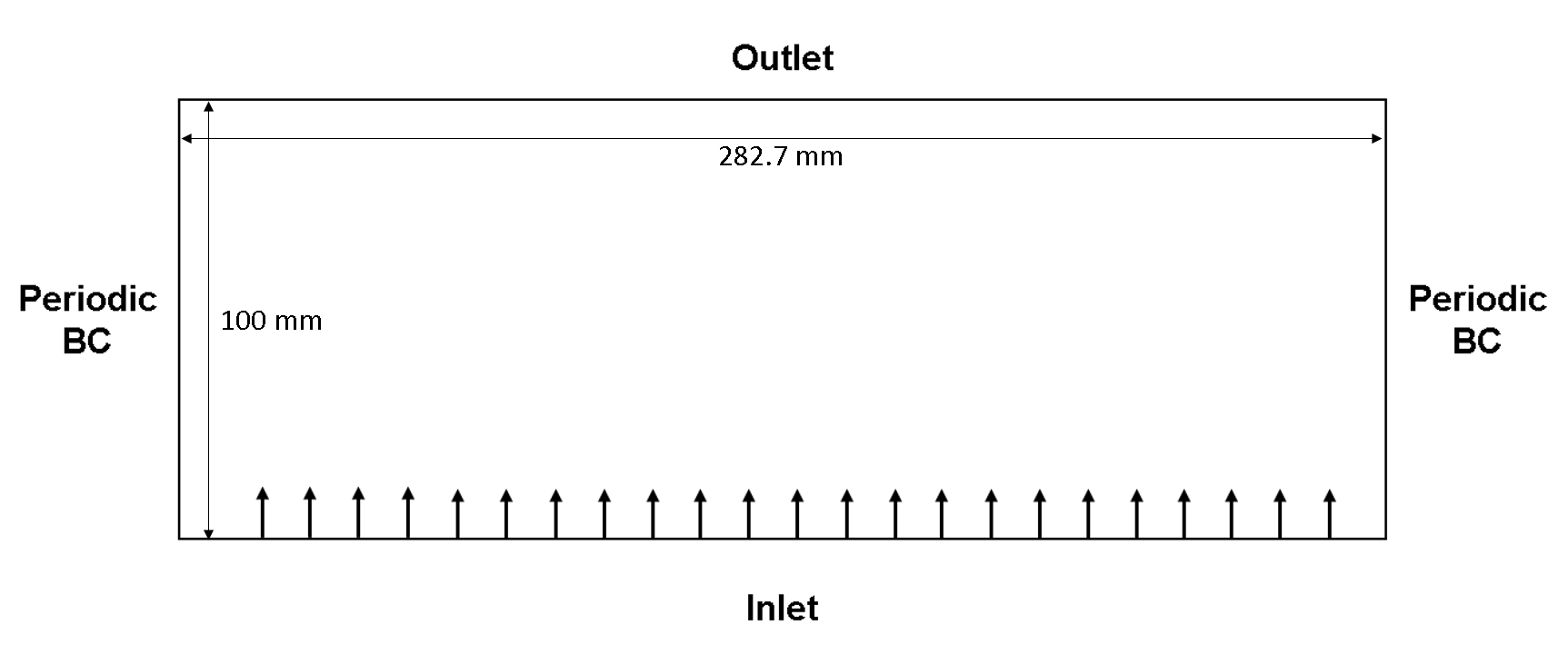}
\caption{Computational domain of the 2D RDE simulation.}\label{FOMsim_fig}
\end{figure}

Following the earlier work by Schwer et al.~\citep{Schwer_2014}, an unrolled two-dimensional (2D) configuration is established to approximate the three-dimensional (3D) RDE geometry as shown in Fig.~\ref{FOMsim_fig}. Both the FOM and ROM calculations are performed using this 2D configuration in the current study. The computational domain has a dimension of 282.7 mm in the azimuthal direction with 1414 uniformly distributed cells ($\Delta x = 0.2\: \text{mm}$) and 100 mm in the axial direction with 125 exponentially distributed cells ($\Delta y_{\text{min}} = 0.2\: \text{mm}$ and $\Delta y_{\text{max}} = .8\: \text{mm}$). This determination of cell size was done to match the resolution used in the Schwer configuration Periodic boundary conditions are imposed on the left and right side of the domain in order to simulate the traveling of the detonation wave around the annulus as in the 3D configuration. Following similar approach by Schwer and Kailasanath ~\citep{Schwer_2011}, to accommodate for the effects of the inlet injection plane from a RDE, a variable inflow condition is applied at the inlet based on the injector conditions $(p,T,v)$ computed assuming isentropic expansion through the injector nozzle into the combustion chamber 
\begin{equation}
    \begin{cases}
        \text{if $p > p_{\text{cr}}$, no flow at the inlet and boundary is treated as a wall} \\
        \text{else, constant velocity, $V_{\text{in}}$, and temperature, $T_{\text{in}}$, are applied at the inlet}
    \end{cases}
\end{equation}
where $T_{\text{in}}$ is set to be 300 K and the critical pressure, $p_{\text{cr}}$, is defined in terms of the stagnation pressure and reactant specific heat ratio, $\gamma_R$
\begin{equation}
    p_{\text{cr}} = p_0 \left( \frac{2}{\gamma_R + 1} \right)^\frac{\gamma_R}{\gamma_R - 1}
\end{equation}
with $p_0$ as the stagnation pressure. A back pressure of 1 atm is applied at the outlet of the computational domain. The computational infrastructure used for the FOM and ROM solves conservation equations for mass, momentum, energy and species transport in a fully coupled way using the in-house CFD code, the General Mesh and Equation Solver (GEMS), which has been used to model a variety of complex, practical reacting flow problems ~\citep{Harvazinski_2015,Huang_2020}. More details of the FOM equations can be found in Appendix ~\ref{appendix:fom_eq}. Numerical simulations are performed at five different inlet velocities, $V_{\text{in}}$ = 100, 125, 150, 175, and 200 m/s, and are used as the test cases to assess the ROM's capabilities in predicting dynamics changes due to parametric variations. 
The lower bound of this range (100 m/s) was selected as lower inlet velocity conditions could not sustain a stable detonation wave. The upper bound (200 m/s) was chosen to ensure that the inlet condition remains subsonic in order to mimic real conditions. The intermediate values were chosen in order to allow for equal intervals between these cases. As will be seen in Section ~\ref{sec:results}, there is not a staggering variation in the resulting detonation waves, which should aid the performance of the different ROMs being tested.

The combustion process in the 2D RDE problem is modeled using a simplified single-step reaction, two-species reaction 
($\textbf{R} \rightarrow \textbf{P}$), which treats both the reactant (\textbf{R}) and product (\textbf{P}) species as calorically perfect gases with identical molecular weights. The simplified reaction model is used in the current study to make the 2D RDE model more accessible to a broader community of researchers in data-driven model development, while maintaining the challenging physics. Here, the thermal properties of the reactant \textbf{R} and the product \textbf{P} are selected to represent a hydrogen-air mixture with equivalence ratio of one and the combustion products (nitrogen oxides and water vapor) to mimic the chemical reaction model in the work by Schwer and Kailasanath~\citep{Schwer_2011}. The chemical reaction source term for the reactant in Eq.~\ref{eq:fom:source_term}  follows in the Arrhenius form 
\begin{equation}
    \dot{\omega}_\text{Reactant} = -MW_\text{Reactant} \cdot A \text{exp} \left( \frac{-E_A/R_u}{T} \right) \left[ \frac{\rho Y_\text{Reactant}}{MW_\text{Reactant}} \right]^a ,
    \label{1Dsource_term}
\end{equation}
with the pre-exponential factor $A = 1.6 \times 10^{11}$, the activation energy $E_A/R_u = 24{,}358$ K, and the concentration exponent $a = 1.0$, which is designed to model the conversion of perfectly-mixed reactants to completely-burned products to generate a detonation wave at the nominal flow conditions. All the numerical simulations are computed using the second-order accurate backwards differentiation formula with dual time-stepping, and a constant physical time step size of $\dt = 2\: $ns using 10 compute nodes (Dual Intel Xeon 6442Y CPUs) with 256GB 4800MT/s DDR5 memory.

\section{Numerical Results and Analysis}
\label{sec:results}

\begin{figure}[hbt!]
    \centering
    \begin{subfigure}[b]{0.6\textwidth}
         \centering
         \includegraphics[width=10cm,trim={4.25cm 1cm 4cm 1cm},clip]{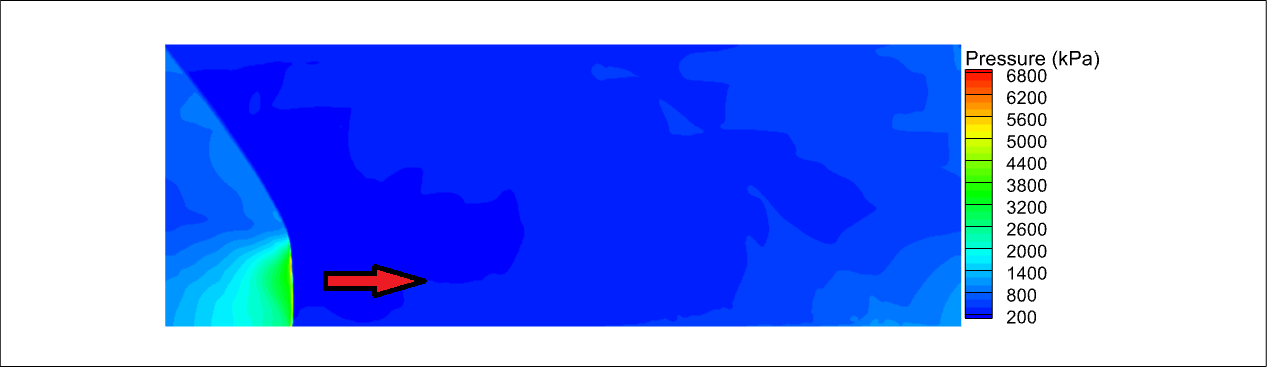}
     \end{subfigure}
     \hfill
     \begin{subfigure}[b]{0.6\textwidth}
         \centering
         \includegraphics[width=10.4cm,trim={5.75cm 1cm 4cm 1cm},clip]{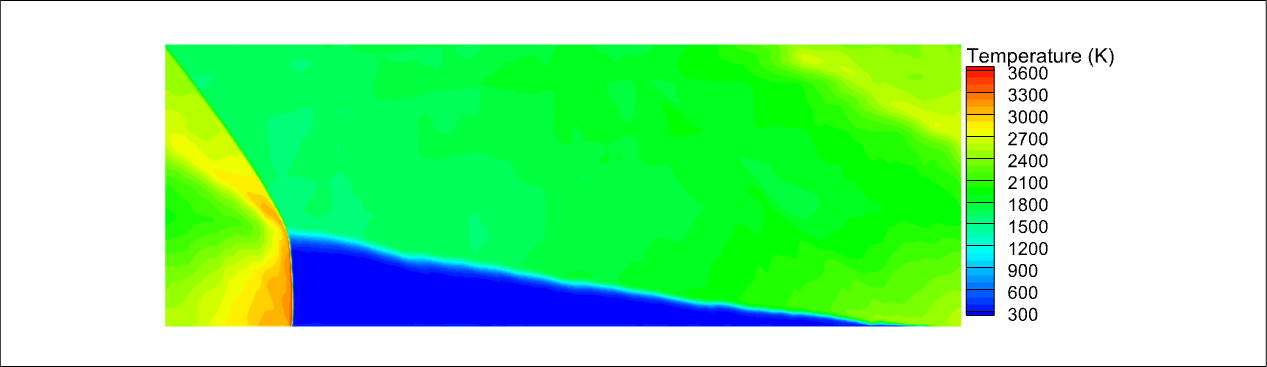}
     \end{subfigure}
        \caption{Representative instantaneous snapshots of pressure (top) and temperature (bottom) from FOM simulation of the 2D RDE at $t = 1 \ \text{ms}$.}
        \label{fig:FOM_Results}
\end{figure}

\begin{figure}[hbt!]
    \centering
    \begin{subfigure}[b]{.75\textwidth}
        \centering
    \includegraphics[width=\textwidth]{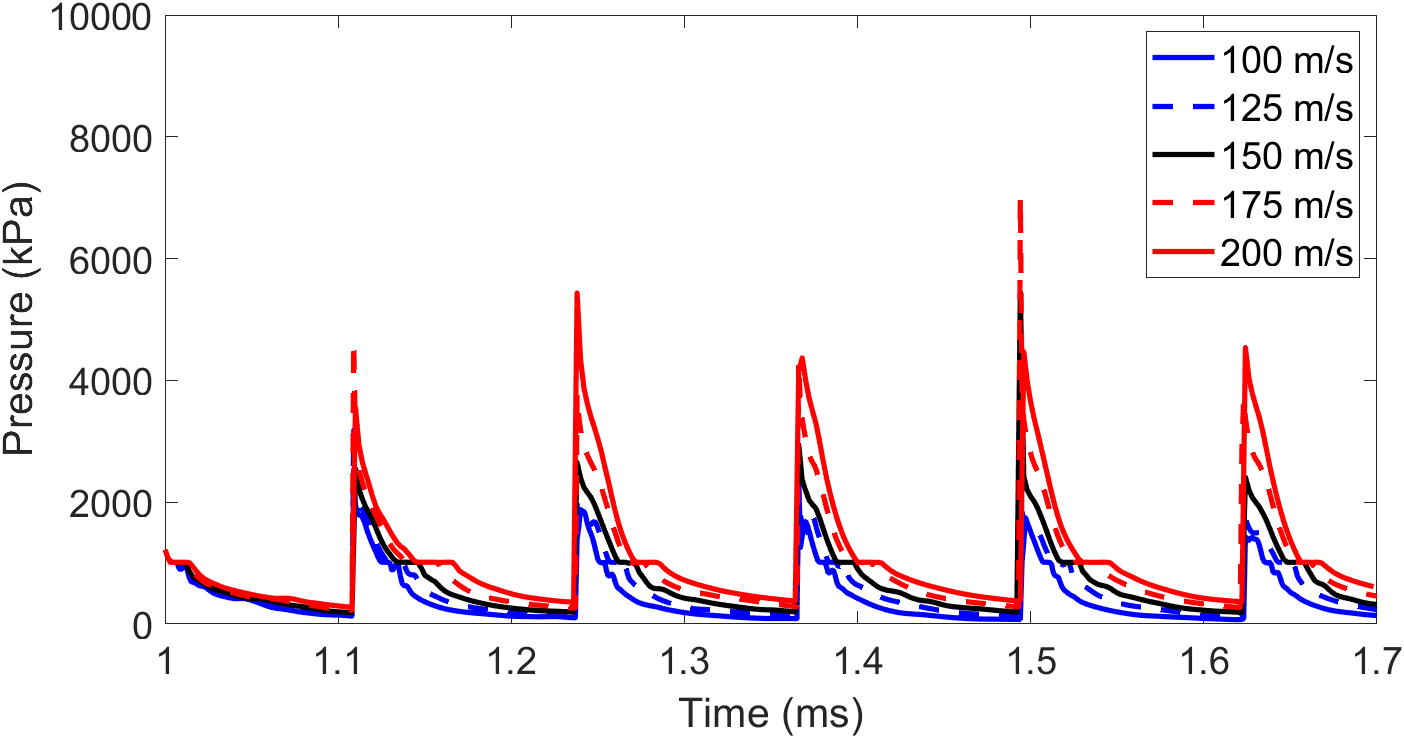}
     \end{subfigure}
     \caption{Local pressure trace results for FOM test cases.}
     \label{fig:FOM_Pressure}
\end{figure}

\begin{figure}[hbt!]
    \centering
    \begin{subfigure}[b]{.4\textwidth}
        \centering
        \captionsetup{oneside,margin={-.75in,0cm}}
        \moveleft .4in \hbox{\includegraphics[width=\textwidth]{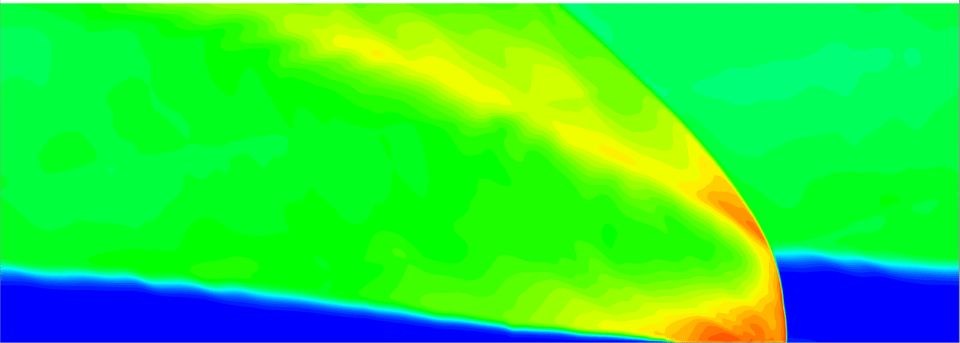}}
        \caption{$V_{\text{in}} =$ 100 \text{m/s}}
        \label{fig:100msFOM}
    \end{subfigure}
    \begin{subfigure}[b]{.4\textwidth}
        \centering
        \captionsetup{oneside,margin={-.75in,0cm}}
        \moveleft .4in \hbox{\includegraphics[width=\textwidth]{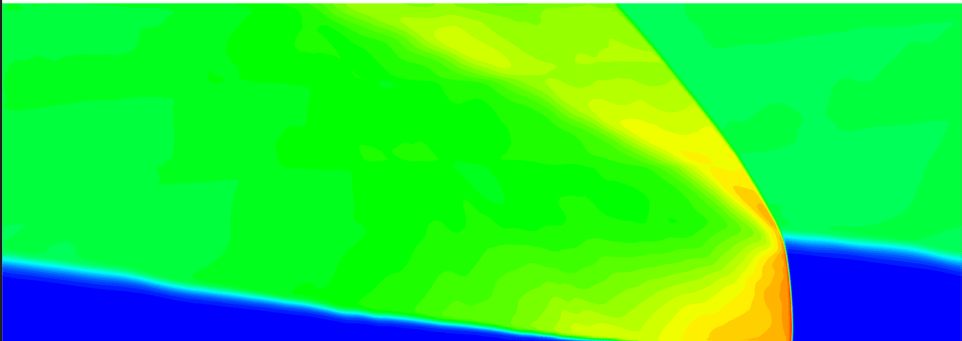}}
        \caption{$V_{\text{in}} =$ 125 \text{m/s}}
        \label{fig:125msFOM}
    \end{subfigure}
    \begin{subfigure}[b]{.4\textwidth}
        \centering
        \includegraphics[width=\textwidth]{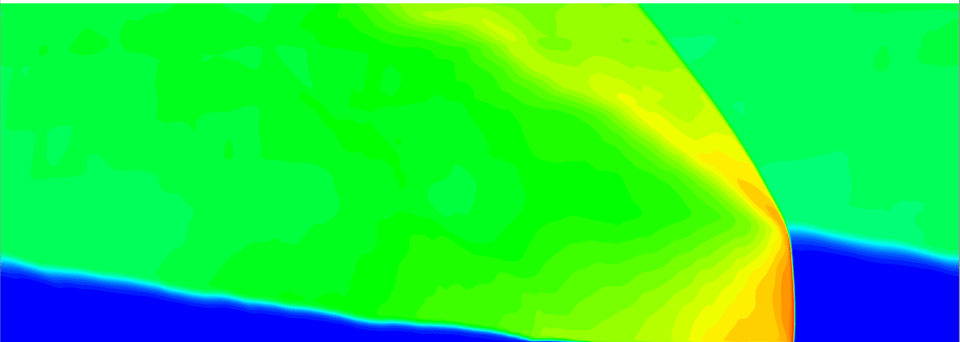}
        \caption{$V_{\text{in}} =$ 150 \text{m/s}}
        \label{fig:150msFOM}
    \end{subfigure}
    \begin{subfigure}[b]{.4\textwidth}
        \centering
        \includegraphics[width=\textwidth]{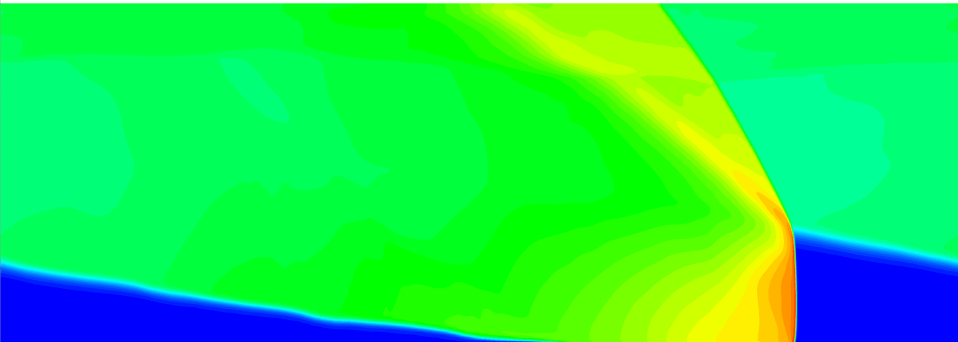}
        \caption{$V_{in} =$ 175 \text{m/s}}
        \label{fig:175msFOM}
    \end{subfigure}
    \begin{subfigure}[b]{.12\textwidth}
        \centering
        \raisebox{-0.05\height}{\includegraphics[width=\textwidth]{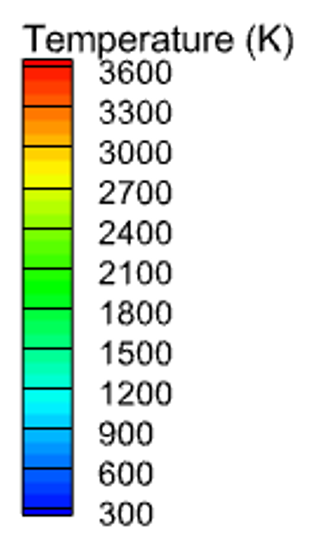}}
    \end{subfigure}
    \begin{subfigure}[b]{.4\textwidth}
        \centering
        \captionsetup{oneside,margin={-.75in,0cm}}
        \moveleft .4in \hbox{\includegraphics[width=\textwidth]{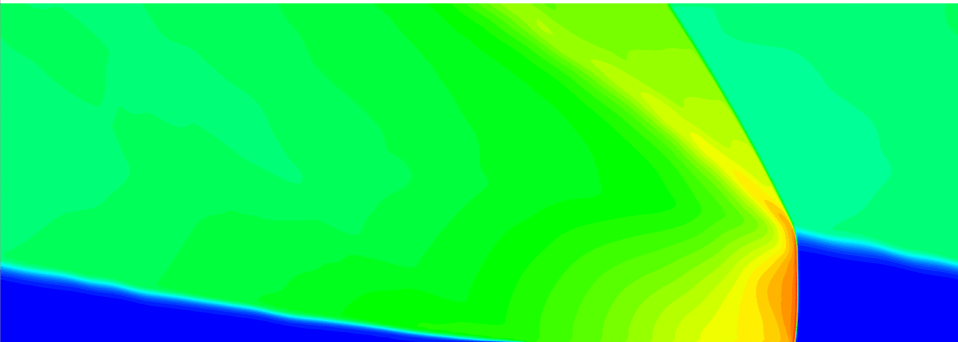}}
        \caption{$V_{\text{in}} =$ 200 \text{m/s}}
        \label{fig:200msFOM}
    \end{subfigure}
        \caption{Representative instantaneous snapshots of temperature from FOM simulation of the 2D RDE at $t = 1.6 \ \text{ms}$.}
        \label{fig:FOM_Results_End}
\end{figure}

The baseline unsteady FOM simulation of the 2D RDE configuration is first conducted at $V_{\text{in}} = 150\: \text{m/s}$, advancing from 0 to 1 ms to establish a stable rotating detonation wave in the domain as shown in Figure \ref{fig:FOM_Results}. Initialized from the solution at 1 ms with $V_{\text{in}} = 150\: \text{m/s}$, a parametric dataset is then generated by conducting FOM simulations at five different equally spaced inlet velocities ($V_{\text{in}} =$ 100, 125, 150, 175, 200 m/s), which leads to distinct detonation-wave dynamics. Figure~\ref{fig:FOM_Pressure} compares the local pressure traces at different inlet velocities, which are measured at the bottom-left corner of the domain in Fig.~\ref{FOMsim_fig} and show clear lifts in pressure levels as the inlet velocity increases, indicating the amplification of the detonation waves as more reactant is injected into the domain. The effects of inlet velocity on detonation-wave dynamics are also reflected by comparing the instantaneous temperature contours as shown in Fig.~\ref{fig:FOM_Results_End}, which exhibits noticeable weakening of the detonation waves with the decrease of inlet velocity, resulting in a shorter and more curved wave front. A decrease in the length of the contact surface is observed on the inlet side of the wave and a decrease in the inclination of the oblique shock and shear layer are observed as well. This parametric dataset establishes an ideal test bed to evaluate the ROM performance in modeling RDE dynamics.

Moreover, due to the fact that all simulations are initialized from the 150 \text{m/s} case at 1 ms, this dataset also contains transient dynamics at the initial establishment of the detonation waves at the corresponding conditions and thus presents additional challenges for ROM development. We remark that this dataset is purposely configured to mimic the possible flow-condition variations during RDE operations, and is used to evaluate the capabilities of the nonlinear quadratic basis and adaptive ROM in capturing RDE dynamics under practical parametric variations. The solution snapshots from 1 to 1.7 ms are collected and used for all the numerical analyses and evaluations in the current paper, which spans a total of 5 cycles (5 complete annular motions of the detonation waves) and each cycle contains 7,000 snapshots.  

\subsection{POD Characteristics}

\begin{figure}[hbt!]
    \centering
    \includegraphics[width=.6\textwidth]{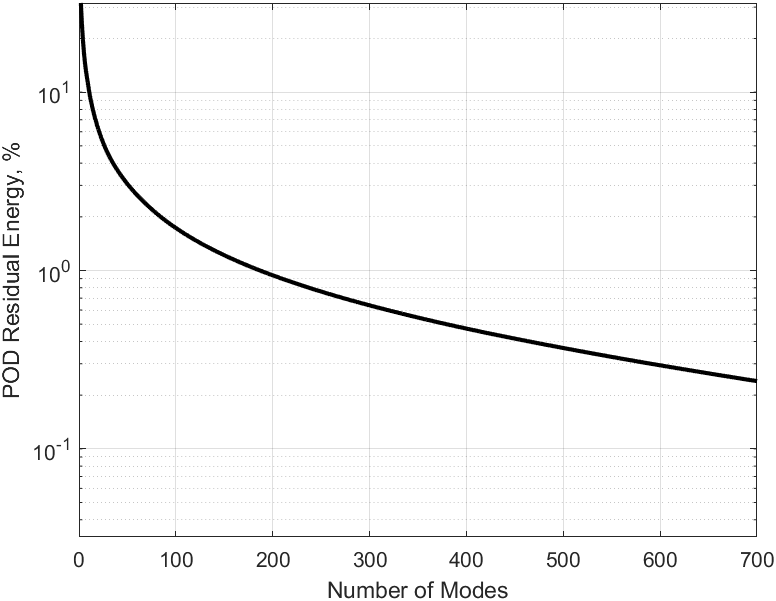}
    \caption{POD Residual energy versus the number of POD modes included.}
    \label{fig:ResidualEnergyPara}
\end{figure}

The POD characteristics of the 2D RDE are first investigated to understand how well the POD trial basis represents the detonation wave dynamics in FOM dataset as shown in Fig.~\ref{fig:ResidualEnergyPara}, which is evaluated using the POD residual energy
\begin{equation}
    \text{POD Residual Energy}(\numSolModes), \;\% = \lp 1 - \frac{\sum^{\numSolModes}_{\genIdx=1} \sigma^2_\genIdx}{\sum^{\numSolModesTotal}_{\genIdx=1} \sigma^2_\genIdx} \rp \times 100,
    \label{eq:pod:res_energy}
\end{equation}
where $\sigma_\genIdx$ is the $\genIdx^\text{th}$ singular value used to compute the trial basis \trialBasisPrim. The singular values are arranged in descending order. $\numSolModes$ is the number of vectors retained in the POD trial basis, and $\numSolModesTotal$ is the total number of modes in the training dataset. An in depth \textit{a priori} projection analysis of the performance of linear static basis for different training data sets and included mode amounts can be found in Appendix \ref{appendix:training}. The results of this analysis informed the decision to utilize a training set composed of 1 cycle of data (7000 time snapshots) each for $V_{\text{in}}$ = 100, 150, and 200 \text{m/s} (21,000 time snapshots total). Figure ~\ref{fig:ResidualEnergyPara} highlights the problem of slow Kolmogorov N-width decay, illustrating that as further accuracy is desired for this convection and shock dominated problem, many more POD modes will need to be included in the trial basis. For example, it can be seen that in order to capture 97\% of the POD residual energy, 40 modes are needed. However, in order to capture 99\% of the residual energy, 200 modes are required, resulting in a much larger POD basis and, therefore, a much more computationally expensive ROM. While in order to reach 99.9\% of the residual energy (a common practice in the literature~\cite{Carlberg2017}), at least 1,180 modes are required, which becomes unaffordable for efficient ROM construction.

\subsection{Assessment of Linear and Nonlinear Quadratic Bases}
\label{sec:apriori}
First, we evaluate and compare the performance of the linear and nonlinear quadratic bases in representing the detonation-wave dynamics via an \textit{a priori} analysis by computing the projected FOM solution($\Bar{\mathbf{q}}^{n}_{p}$ and $\Bar{\mathbf{q}}^{n}_{p,\text{quadratic}}$)
\begin{equation}
    \Bar{\mathbf{q}}_p = \mathbf{q}_{p,\text{ref}} + \scaleMatPrim^{-1} \trialBasisPrim \mathbf{q}_r .
    \label{eq:projFOML}
\end{equation}
\begin{equation}
    \begin{aligned}
    \Bar{\mathbf{q}}^{n}_{p,\text{quadratic}} & = \mathbf{q}_{p,\text{ref}} + \scaleMatPrim^{-1} \left[ \trialBasisPrim \hat{\mathbf{q}}^n_r + \Bar{\mathbf{V}}_{p}(\hat{\mathbf{q}}^n_r \otimes \hat{\mathbf{q}}^n_r) \right]
    \end{aligned}
    \label{eq:projFOMNL}
\end{equation}
where $\hat{\mathbf{q}}^n_r=\trialBasisPrim^T \left( \mathbf{q}^n_p - \mathbf{q}_{p,\text{ref}} \right)$ and $\mathbf{q}^n_p$ corresponds to the state solution obtained at the $n^{th}$ FOM snapshot. It is noteworthy that the projected FOM solution determines the upper bound of the resulting ROMs' accuracy (i.e., the accuracy of the ROM solution cannot outperform the corresponding projected FOM solution). This \textit{a priori} analysis based on projected FOM solutions has been demonstrated as an effective approach to measure the feasibility and expected performance of constructed trial basis in representing target dynamics of interest without the need to develop an online ROM~\citep{HuangMPLSVT2022,Arnold-Medalbalimi_2022,McQuarrieOpInf2021}. 

\begin{figure}
    \centering
    \includegraphics[width=.6\textwidth]{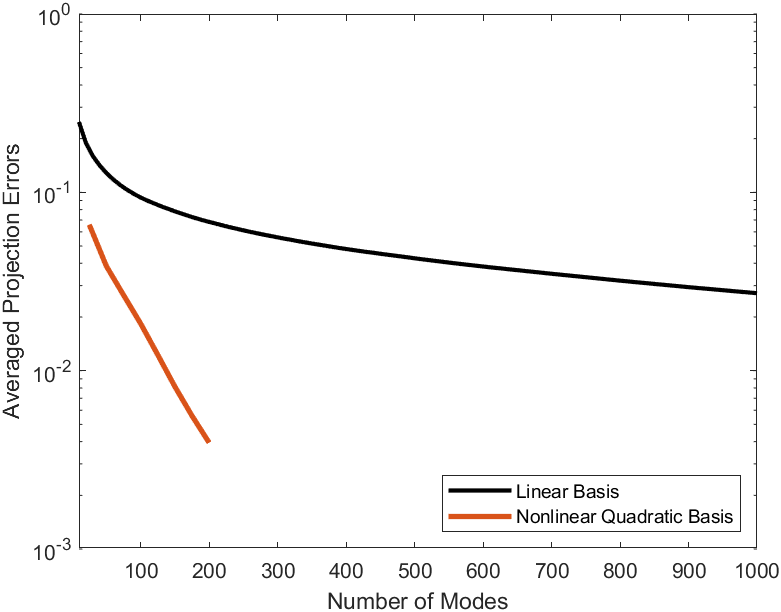}
    \caption{Comparisons of the projection errors between linear static and quadratic bases in reproducing the training dataset.}
    \label{fig:NLErrorComparisonUp}
\end{figure}

As mentioned previously, one cycle of FOM snapshots are collected at three representative inlet-velocity conditions ($V_{\text{in}} =$ 100, 150, and 200 \text{m/s}), following the findings in Appendix\ref{appendix:training} for the most efficient training data set. This resulted in a total of 21,000 training snapshots, taken at every 10 time step of the simulation. Solutions at the other two conditions (i.e., $V_{\text{in}} =$ 125 and 175 \text{m/s}) are used as the testing snapshots to evaluate the capabilities of the linear and nonlinear quadratic bases in representing the changes in dynamics due to parametric variations. Specifically, the projection errors, $\errProjVsFOM^{\iterIdx}$, are quantified for the evaluations
\begin{equation}
    \errProjVsFOM^{\iterIdx} = \frac{1}{\numVars} \sum^{\numVars}_{\varIdx=1} \frac{\norm{\solPrimROMProjVar^{\iterIdx} - \solPrimFOMVar^{\iterIdx}}}{\norm{\solPrimFOMVar^{\iterIdx}}} ,
    \label{eq:pod:proj_err}
\end{equation}
where $\numVars$ represents the number of included quantities of interest (i.e. pressure, temperature), $\solPrimROMProjVar^{\iterIdx}$ represents the $i^\text{th}$ solution variable of the projected FOM state vector, $\solPrimROMProj^{\iterIdx}$, at time step $\iterIdx$. The projection errors are computed for each training condition ($V_{in}$) within the training FOM snapshots using both linear and nonlinear quadratic bases with different numbers of modes included, which are compared in Fig.~\ref{fig:NLErrorComparisonUp} for $V_{\text{in}} =$ 100 \text{m/s}. It can readily seen that the nonlinear quadratic basis provides one order of magnitude improvement in the projection error in contrast to the linear basis with lower dimensionality. We remark that different scalar regularization parameters, $\lambda$, in Eq.~\ref{eq:nonlinear_vbar_reg} are tested to compute the nonlinear quadratic basis based on the projection errors and the results of  are included in Appendix \ref{appendix:regularization}. A value of $\lambda = 10^{-4}$ is shown to be sufficient to ensure a well-conditioned least-square problem in Eq.~\ref{eq:nonlinear_vbar} and is used to construct the nonlinear quadratic basis in the current study.  The maximum number of modes included in the \textit{a priori} analysis is set to be 200 as a result of the significantly increased dimension of $\Bar{\mathbf{V}}_{p}$, which is $n^2_p$. Though by eliminating
redundancy in the Kronecker products in Eq.~\ref{eq:nonlinear_vbar}, the dimension of $\Bar{\mathbf{V}}_{p}$ can be reduced to $n_p(n_p+1)/2$, for a linear basis with $n_p = 200$, the resulting nonlinear quadratic basis, $\Bar{\mathbf{V}}_{p}$, would be of dimension 20,100. It should be noted that while the dimensionality of $\Bar{\mathbf{V}}_{p}$ increases as shown with the number of included modes, the dimensionality of the resulting ROM remains the same. Such a large dimension for $\Bar{\mathbf{V}}_{p}$, however, requires significant computational resources, especially in basis computing and storage. Therefore, though increasing the dimension of $\Bar{\mathbf{V}}_{p}$ is likely to produce more accurate results, the associated computational expense of generating the corresponding nonlinear quadratic basis can quickly become untenable.

\begin{figure}
    \centering
    \begin{subfigure}[b]{.7\textwidth}
        \centering
    \includegraphics[width=0.85\textwidth]{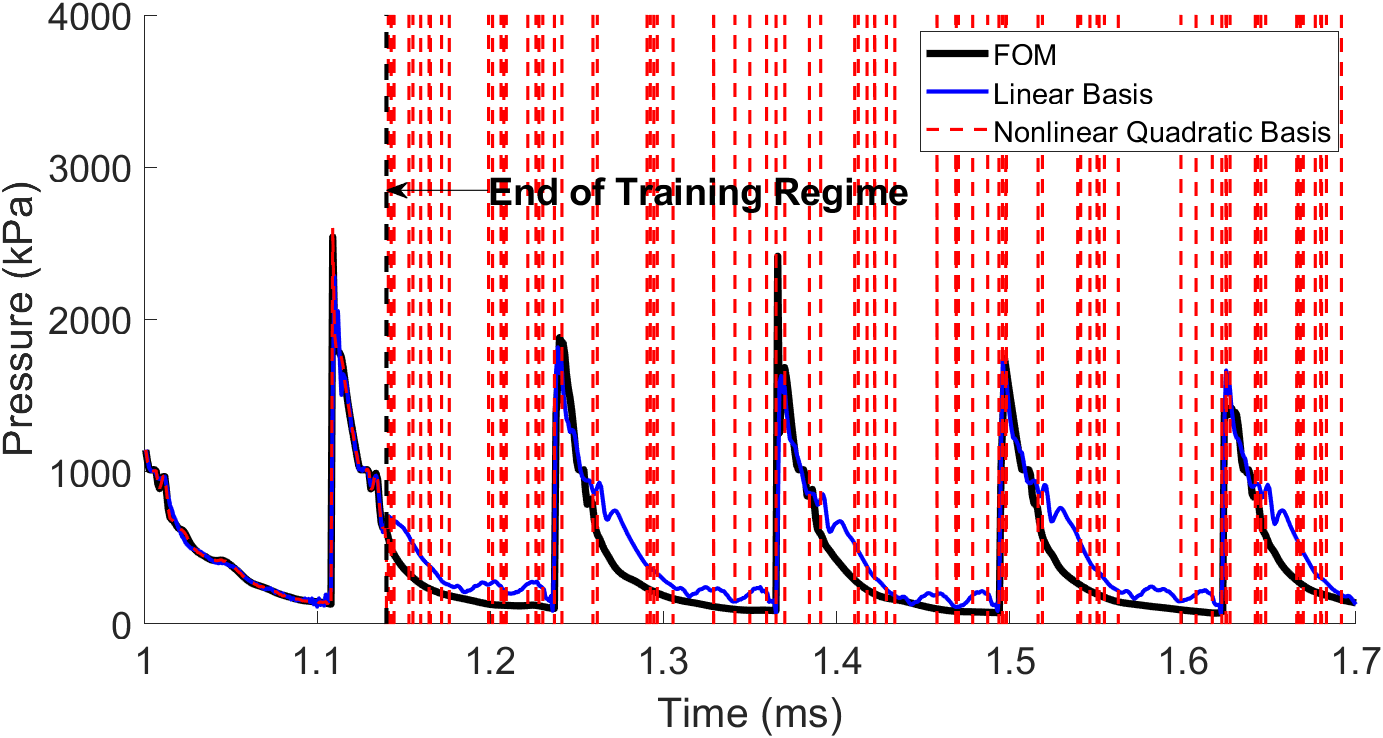}
        \label{fig:100Press}
    \end{subfigure}
    \caption{Comparisons of local pressure time traces between the FOM and the projected FOM solutions using linear and nonlinear quadratic  bases with 200 modes included at $V_{\text{in}} =$ 100 \text{m/s}.}
    \label{PressureMonitorNonlinear_100}
\end{figure}

\begin{figure}
    \centering
    \begin{subfigure}[b]{.471\textwidth}
        \centering
        \captionsetup{oneside,margin={.85in,0cm}}
        \includegraphics[width=\textwidth]{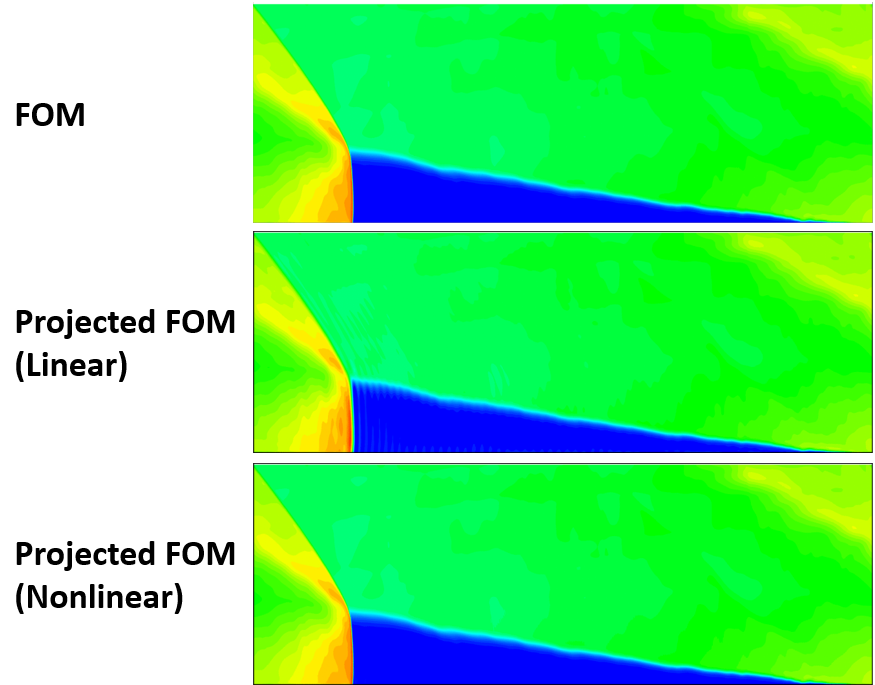}
        \caption{$t =$ 1.125 \text{ms}}
        \label{fig:100CompT}
    \end{subfigure}
    \begin{subfigure}[b]{.340\textwidth}
        \centering
        \includegraphics[width=\textwidth]{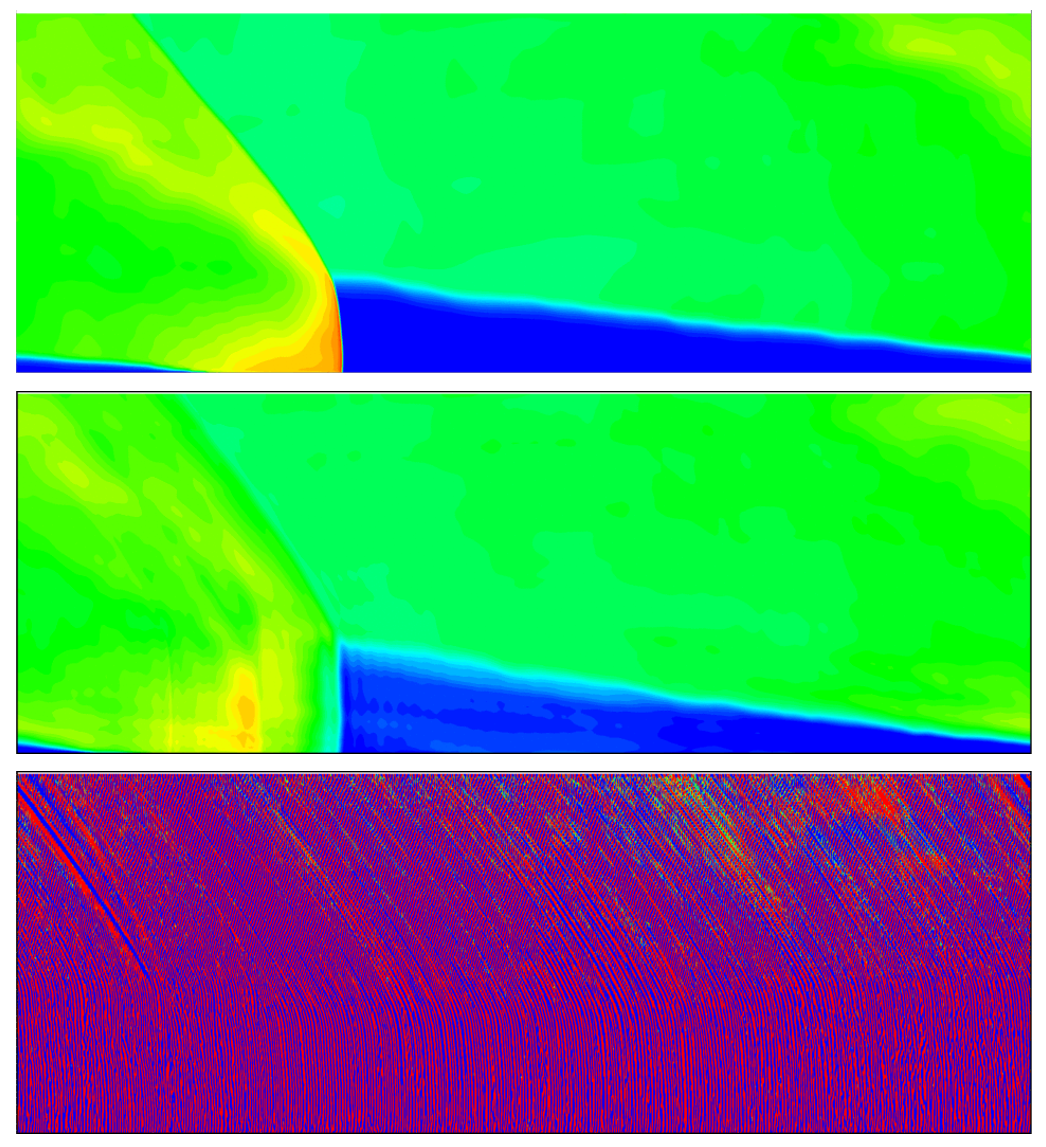}
        \caption{$t =$ 1.150 \text{ms}}
        \label{fig:100CompOT}
    \end{subfigure}
    \centering
    \begin{subfigure}[b]{.15\textwidth}
        \centering
        \raisebox{0.5\height}{\includegraphics[width=\textwidth]{figures/ColorBar.JPG}}
    \end{subfigure}
    \caption{Comparisons of instantaneous temperature fields between the FOM and the projected FOM solutions using linear and nonlinear quadratic bases with 200 modes included at $V_{\text{in}} =$ 100\ \text{m/s}.}
    \label{100msPredictionComparison}
\end{figure}

Next, we proceed to examine and compare the local pressure time traces and instantaneous field results between the FOM and the projected FOM solutions to assess the capabilities of the nonlinear bases in representing the detonation wave dynamics at the training conditions ($V_{\text{in}} =$ 100 \text{m/s}). First, The local pressure time traces are compared in Fig.~\ref{PressureMonitorNonlinear_100} using 200 modes to construct the bases at a probe placed at the same location as in Fig.~\ref{fig:FOM_Pressure}, which shows that the projected FOM solutions using the nonlinear quadratic basis closely recreate the FOM results within the training regime. However, it can be readily seen from that the nonlinear quadratic basis fails to provide physical representations of the future-state dynamics immediately outside the training regime. On the other hand, though the linear basis leads to less accurate projected solutions within the training regime, it provides reasonably accurate representation of the pressure trace in the future state. Second, the instantaneous temperature fields are compared between the FOM and projected FOM solutions in Fig.~\ref{100msPredictionComparison}. At the time instance within the training regime ($t=$ 1.125 \text{ms}), consistent with Fig.~\ref{PressureMonitorNonlinear_100}, the projected FOM solutions using the nonlinear basis reaches better agreement with the FOM while spurious oscillations are present in the linear-basis results. However, neither the linear or the nonlinear basis can provide accurate representations of the temperature fields beyond the training ($t=$ 1.150 \text{ms}) though the local pressure trace in Fig.~\ref{PressureMonitorNonlinear_100} seems to indicate that it is feasible to the linear basis to represent the future-state detonation-wave dynamics. In addition, we remark that the same investigations are performed for $V_{\text{in}} =$ 150 and 200 \text{m/s} which exhibit similar comparisons as for $V_{\text{in}} =$ 100 \text{m/s} in Figs.~\ref{PressureMonitorNonlinear_100} and~\ref{100msPredictionComparison}.  

Furthermore, we continue to evaluate the capabilities of the linear and nonlinear quadratic bases in representing changes in dynamics due to parametric variations by computing the projected FOM solutions at the testing conditions (i.e., $V_{\text{in}}=$ 125 and 175 \text{m/s}). Following the same investigations above, both local pressure time traces and instantaneous temperature fields are compared. It should be noted that since the snapshot solutions at the two inlet velocities are not included in the training dataset for the bases construction, the projected FOM solutions are direct indicator of the feasibility of using these bases in constructing parametric ROMs either through intrusive~\citep{HuangMPLSVT2022} or non-intrusive~\cite{McQuarrieOpInf2021} approaches. Akin to the results shown in Figs.~\ref{PressureMonitorNonlinear_100} and~\ref{100msPredictionComparison}, the nonlinear quadratic basis shows restricted capabilities in representing dynamics beyond the training dataset and produces non-physical solutions in both local pressure traces and instantaneous temperature fields. Similarly, in the 125 m/s study,
the linear basis again provides a reasonably accurate trend for the local pressure evolution for both inlet velocities, although a bias of over or under estimation based on the case in question is apparent. This again does not tell the full story, however, as the results 
reveal the inability of this method to provide accurate unsteady temperature field results over the entire domain as the simulation progresses through time. As early as .001 ms into the simulation, nonphysical striations in the detonation wave front can be seen, with the fidelity of the detonation wave structure continuing to decay as time progresses.

The \textit{a priori} analyses conducted in the current section shows that the nonlinear quadratic basis is able to significantly improve ROM accuracy in reproducing the detonation-wave dynamics within the training regime compared to the linear basis. However, it exhibits very restricted capabilities in representing the dynamics beyond the training regime (either in the future state or at another parameter). This indicates that the resulting ROM constructed using the nonlinear quadratic basis will likely present limited prediction capabilities for RDE applications. In addition, though more flexible in representing dynamics beyond the training regime, the linear-basis ROMs are anticipate to present similar limitations as the nonlinear-basis ROMs.

\subsection{Assessment of Adaptive ROM}

With the \textit{a priori} analysis showing restrictions in using either the linear or nonlinear quadratic \textit{static} bases to represent RDE dynamics, we then proceed to evaluate the performance of the adaptive ROM in modeling the 2D RDE problem. It should be noted that for the investigations in the current section, an online adaptive ROM is constructed, instead of relying on the \textit{a priori} analysis in the previous section. Following the adaptive ROM formulation in Section~\ref{sec:adaptivity}, the initial basis is computed using POD with 10 training snapshots at the condition $V_{\text{in}} =$ 150\ \text{m/s} with 10 modes preserved to start the adaptive ROM. In contrast to a total of 21,000 snapshots used for the static bases construction, the adaptive ROM formulation significantly reduces the required offline cost in ROM construction. In addition, 8,837 sampling points are selected following Eq.~\ref{eq:adaptiverom:sampling_minimization} for all the adaptive ROM calculations in the current study, which constitute 5 $\%$ of the total cells in the computation domain and are updated every 5 time steps (i.e., $z_s = 5$ in Eq.~\ref{eq:adaptiverom:surrogate_fom_us}). 

\begin{figure}
    \centering
    \begin{subfigure}[b]{.47\textwidth}
        \centering
        \includegraphics[width=\textwidth]{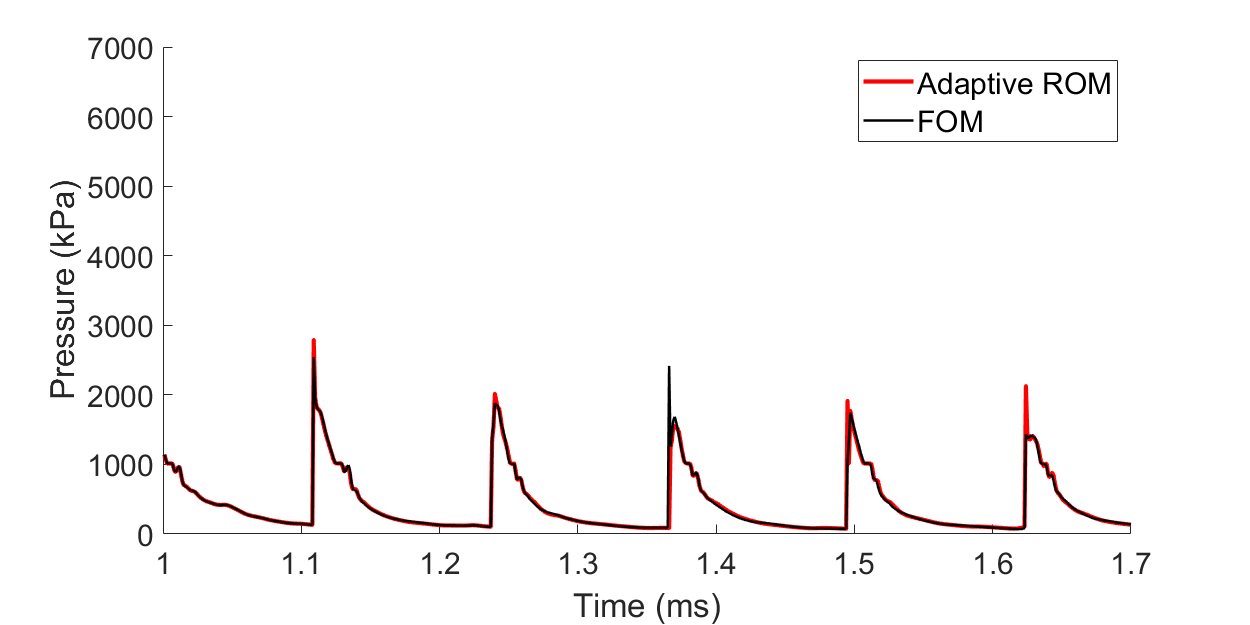}
        \caption{$V_{\text{in}} =$ 100 \text{m/s}}
        \label{fig:100msAdaptiveP}
    \end{subfigure}
    \begin{subfigure}[b]{.4\textwidth}
        \centering
        \includegraphics[width=\textwidth]{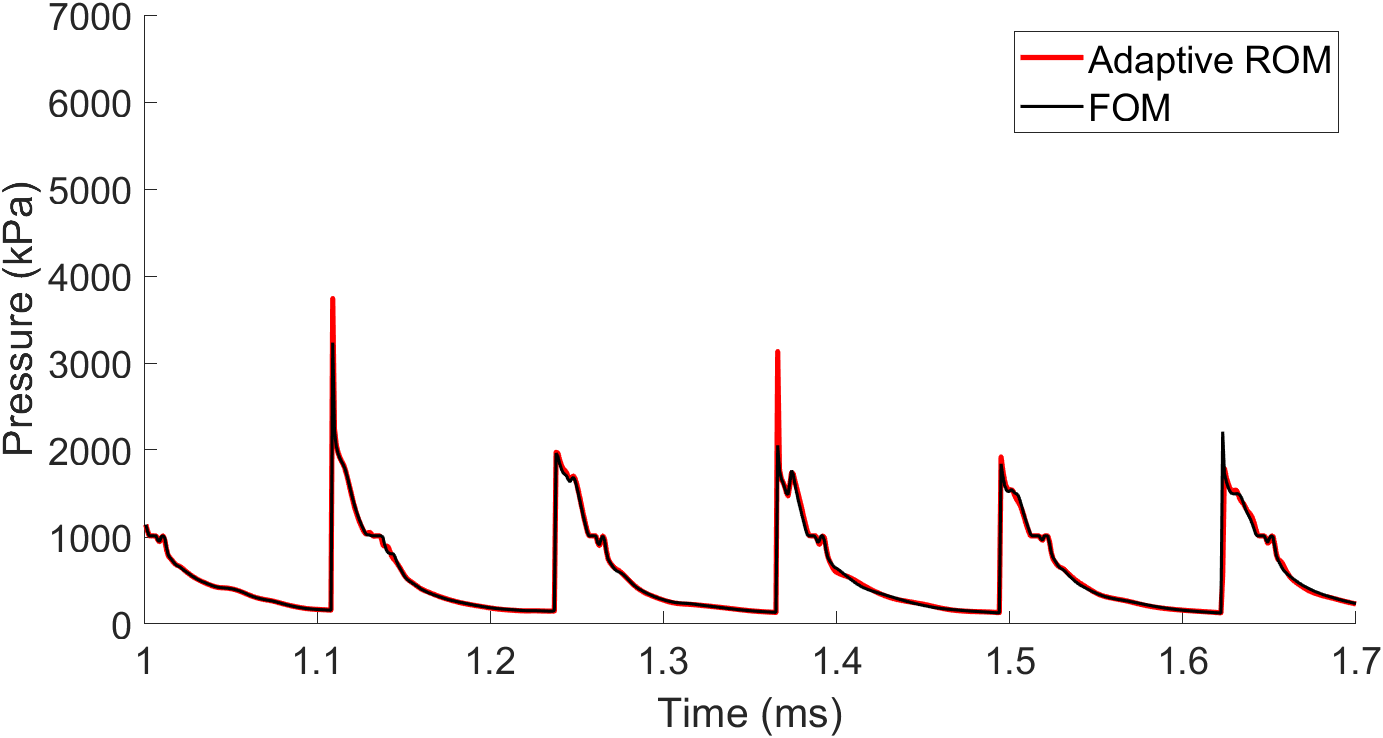}
        \caption{$V_{\text{in}} =$ 125 \text{m/s}}
        \label{fig:125msAdaptiveP}
    \end{subfigure}
    \begin{subfigure}[b]{.47\textwidth}
        \centering
        \includegraphics[width=\textwidth]{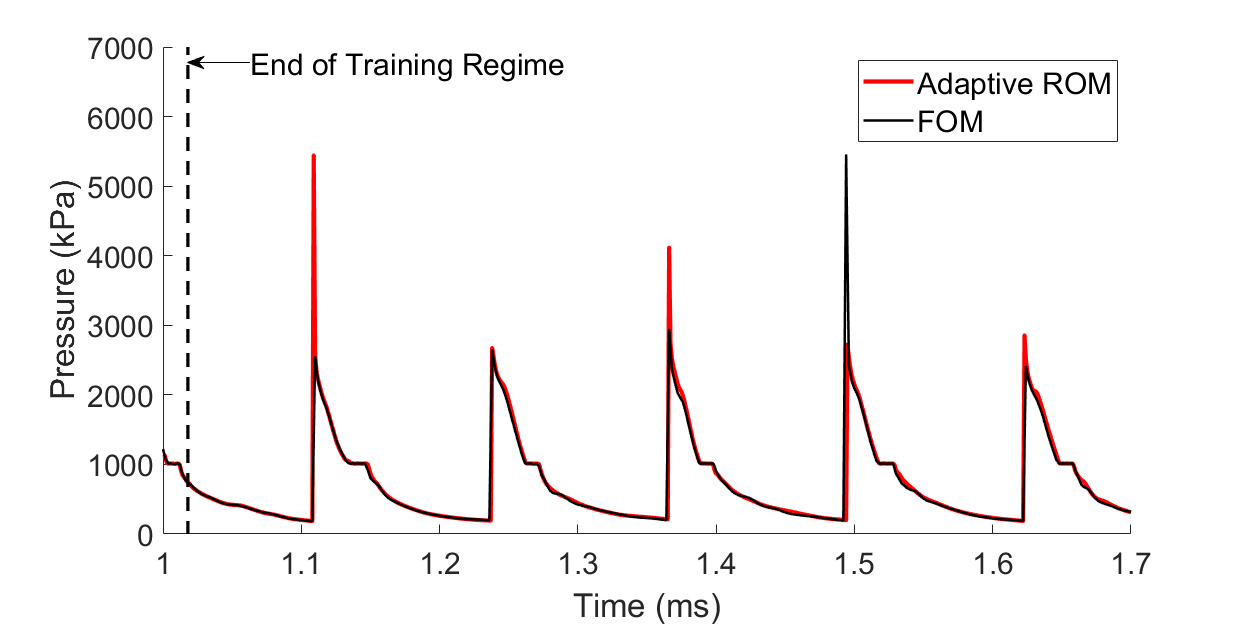}
        \caption{$V_{in} =$ 150 \text{m/s}}
        \label{fig:150msAdaptiveP}
    \end{subfigure}
    \begin{subfigure}[b]{.4\textwidth}
        \centering
        \includegraphics[width=\textwidth]{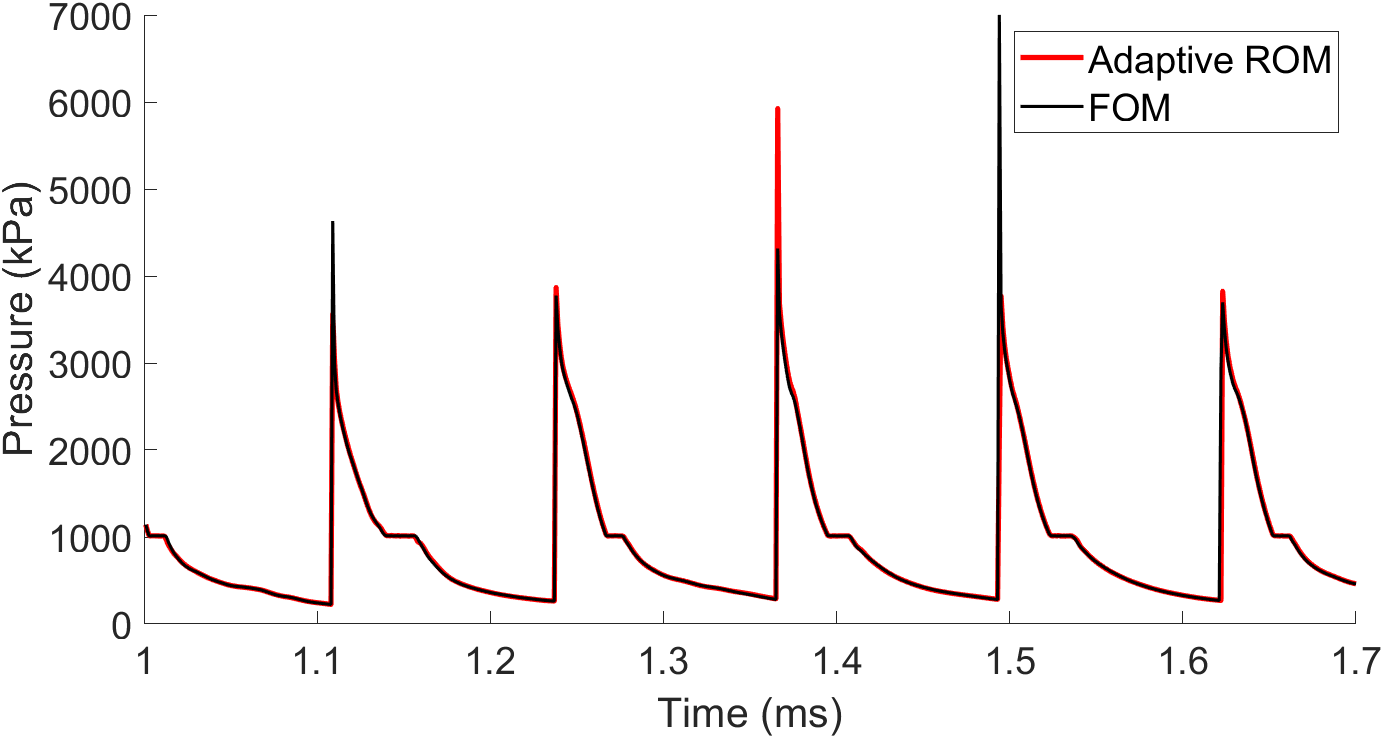}
        \caption{$V_{\text{in}} =$ 175 \text{m/s}}
        \label{fig:175msAdaptiveP}
    \end{subfigure}
    \begin{subfigure}[b]{.47\textwidth}
        \centering
        \includegraphics[width=\textwidth]{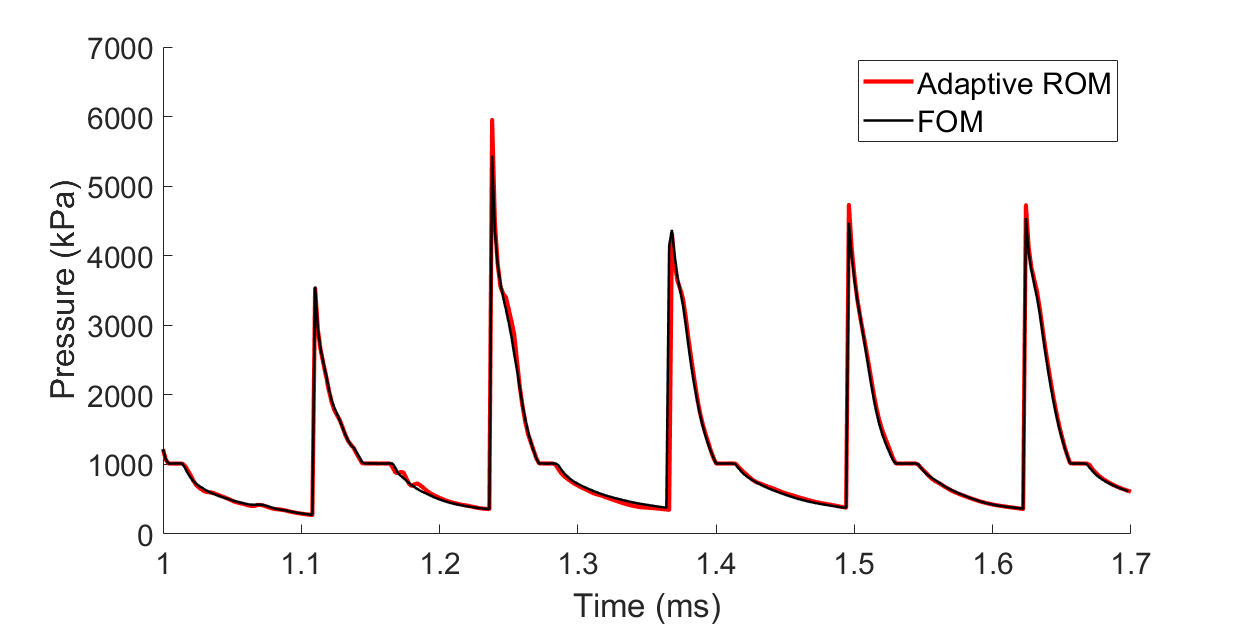}
        \caption{$V_{\text{in}} =$ 200 \text{m/s}}
        \label{fig:200msAdaptiveP}
    \end{subfigure}
    \caption{Comparisons of local pressure time traces between FOM and adaptive ROM at different inlet-velocity conditions.}
    \label{fig:AdaptiveROMPressMon}
\end{figure}

First, the local pressure time traces are compared in Fig.~\ref{fig:AdaptiveROMPressMon} between the FOM and the adaptive ROM at five different inlet velocities and it can be readily seen that the adaptive ROM successfully predicts the changes in pressure dynamics due to parametric variations. Specifically, the adaptive ROM captures the weakening of the detonation-wave intensity, indicated by the peak values of the pressure trace, when the inlet velocity is decreased from 150 \text{m/s} to 125 and 100 \text{m/s}. Moreover, it also accurately predicts the strengthening of the detonation-wave intensity when the inlet velocity is increased from 150 \text{m/s} to 175 and 200 \text{m/s}. More importantly, we remark that the adaptive ROM captures the transient detonation wave dynamics due to the abrupt change of the inlet-velocity at 1 \text{ms}, which is well recognized to be a challenging phenomenon for static-basis ROMs~\citep{HuangMPLSVT2022,Arnold-Medabalimi_2022GTMC}.

\begin{figure}
    \centering
    \begin{subfigure}[b]{.465\textwidth}
        \centering
    \includegraphics[width=\textwidth]{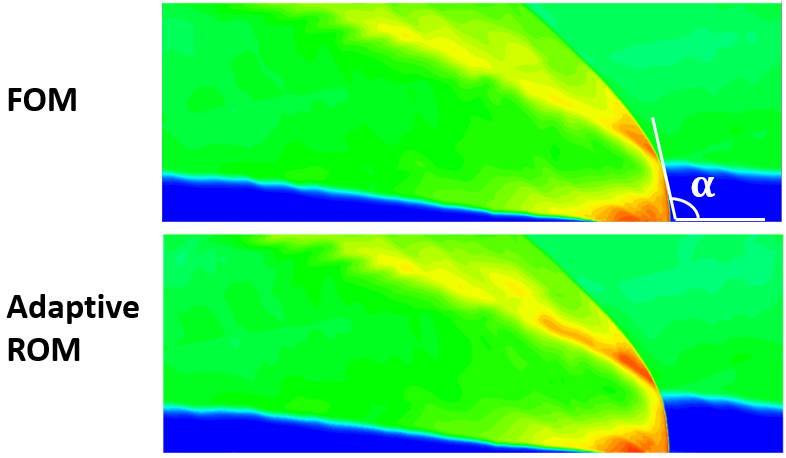}
        \caption{$V_{\text{in}} =$ 100 \text{m/s}}
        \label{fig:100AdapT}
    \end{subfigure}
    \begin{subfigure}[b]{.37\textwidth}
        \centering
    \includegraphics[width=\textwidth]{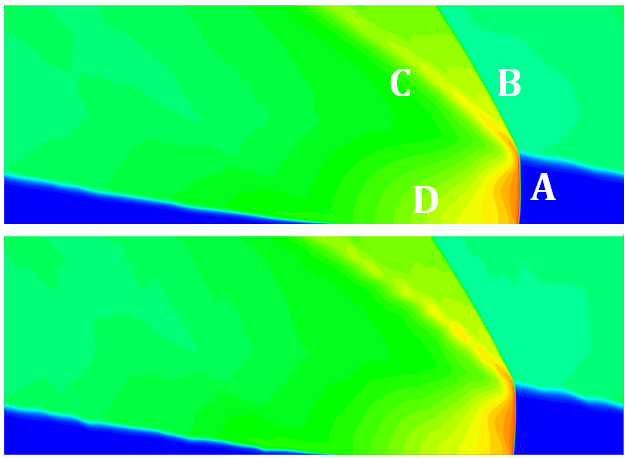}
        \caption{$V_{\text{in}} =$ 200 \text{m/s}}
        \label{fig:200AdapT}
    \end{subfigure}
    \centering
    \begin{subfigure}[b]{.15\textwidth}
        \centering
    \raisebox{0.25\height}{\includegraphics[width=\textwidth]{figures/ColorBar.JPG}}
    \end{subfigure}
    \caption{Comparisons of instantaneous temperature fields between FOM and adaptive ROM at $t =$ 1.6 \text{ms} with relevant features. A) detonation wave, B) oblique shock, C) shear layer, D) contact surface.}
    \label{fig:AdaptivePredictionComparison}
\end{figure}

Next, the instantaneous temperature fields are compared in Fig. \ref{fig:AdaptivePredictionComparison} between the FOM and adaptive ROM at the lowest and highest inlet velocities ($V_{\text{in}} =$ 100 and 200 \text{m/s}) in the testing conditions and a representative time instance near the tend of the testing regime (1.6 \text{ms}) is selected for the comparisons. It can be readily seen that the adaptive ROM solutions reach an excellent agreement with the FOM, especially in capturing the distinct detonation-wave features at different inlet velocities. Though the adaptive ROM overpredicts the temperature in the shear layer at $V_{\text{in}} =$ 100 \text{m/s}, it accurately predicts the characteristics of the detonation waves, oblique shocks, and contact surfaces at both inlet velocities. More importantly, the adaptive ROM successfully predicts the changes in detonation-wave dynamics subject to the increase in the inlet velocity, such as 1) the decrease in the angle ($\alpha$ in Fig.~\ref{fig:AdaptivePredictionComparison}) between the detonation front and the inlet; and 2) the increase inclination in the shear layer and oblique shock. Additionally, when compared to the FOM, the adaptive MOR method demonstrated the capability to provide a factor of 4 computational acceleration. These results, coupled with the computational efficiency gains, highlight the truly predictive capabilities of the adaptive ROM and illustrate its potential to construct ROMs that can be applied to a wide range of operating conditions.

\begin{figure}
    \centering
    \begin{subfigure}[b]{.465\textwidth}
        \centering
        \captionsetup{oneside,margin={.85in,0cm}}
        \includegraphics[width=\textwidth]{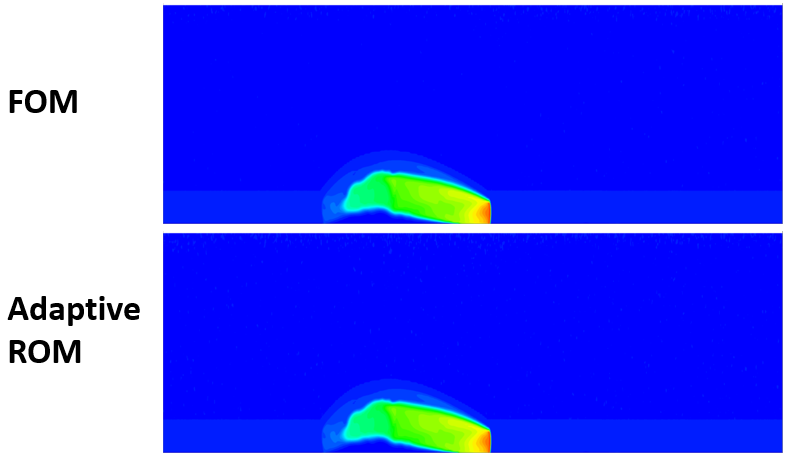}
        \caption{$t =$ .022 \text{ms}}
        \label{fig:22AdapTEarly}
    \end{subfigure}
    \begin{subfigure}[b]{.37\textwidth}
        \centering
        \includegraphics[width=\textwidth]{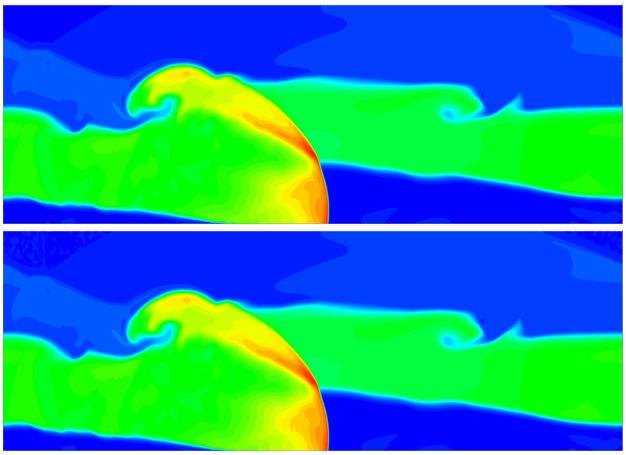}
        \caption{$t =$ .150 \text{ms}}
        \label{fig:150AdapTEarly}
    \end{subfigure}
    \centering
    \begin{subfigure}[b]{.15\textwidth}
        \centering
        \raisebox{0.25\height}{\includegraphics[width=\textwidth]{figures/ColorBar.JPG}}
    \end{subfigure}
    \caption{Comparisons of local pressure trace and instantaneous temperature fields between FOM and adaptive ROM at $V_{\text{in}} =$ 150 \text{m/s}}
    \label{fig:AdaptivePredictionComparisonEarly}
\end{figure}

With the modeling capabilities of adaptive ROM demonstrated for future-state and parametric predictions, we further demonstrate its potential in predicting the initial transience in establishing the rotating detonation waves. As mentioned in Section ~\ref{sec:results}, the FOM simulation has been run for 1 \text{ms} at $V_{\text{in}} =$ 150 \text{m/s} to ensure a stable detonation wave, after which the training data are collected for the investigations presented in the previous sections. While it is necessary to properly assess the performance of the ROM methods based on statistically stationary dynamics, for practice, the computational cost for the FOM simulation of this initial transience (comparable to the testing period from 1 to 1.7 \text{ms}) is fairly significant and the potential cost reduction via ROM for this transience's computation can be crucial. Therefore, we proceed to explore the potential of adaptive ROM to model the initial transience in RDE problems. To achieve this, the FOM is simulated first from 0 to 0.02 \text{ms} (10,000 time steps) to pass over the unphysical solutions due to the artifacts of the initial conditions used to start the simulation. After that, the simulation is switched to adaptive ROM with the initial basis trained using 10 snapshots from 0.02 \text{ms} using the same parameters as in Figs. ~\ref{fig:AdaptiveROMPressMon} and~\ref{fig:AdaptivePredictionComparison}. The resulting local pressure trace and instantaneous temperature fields are compared in Fig.~\ref{fig:AdaptivePredictionComparisonEarly} between the FOM and adaptive ROM. It can be readily seen that the adaptive ROM successfully captures the initial transience in establishing the detonation waves as shown in the FOM, which indicates that it is feasible to apply the adaptive ROM starting at a much earlier stage of the simulations to further reduce the computational cost. 

While the adaptive ROM exhibits great potential and versatility in providing accurate future-state, transient, and parametric predictions for RDE problems with minimal offline training requirements, there remains one aspect that requires further development. The online adaptation of basis and sampling points require additional computation cost to access the FOM equation residual, limiting the computational acceleration that can be achieved via adaptive ROM as highlighted in the work by Huang and Durasaimy~\citep{Huang_2023}. Moreover, scalable implementation of the adaptive ROM algorithm for large-scale engineering problems requires further development, especially in developing an effective load-balancing strategy to accommodate the adapted sampling points, which is currently not incorporated in our adaptive ROM implementation. Therefore, the current adaptive ROM exhibits only $O(4)$ acceleration in computational time compared to the FOM. Huang and Duraisamy have shown, however, that a theoretical acceleration of $> O(20)$ is possible with the adaptive ROM methodology ~\citep{Huang_2023}. It has been shown that with proper dynamic load-balancing, this theoretical limit can be approached quite closely ~\citep{Arnold-Medabalimi_2023}. Therefore a focus of our future work will be to implement such load-balancing for the RDE problem.
\section{Conclusions}
\label{sec:conclusion}
The current work focuses on investigating the capabilities of representative projection-based ROMs constructed utilizing conventional linear static basis, novel nonlinear quadratic basis, and a recently developed adaptive model order reduction (MOR) formulation, in parametric modelling of rotating detonation engines (RDEs). RDEs exhibit strong convection-dominated shock-wave physics, which have been shown to be challenging for ROM development. A two-dimensional (2D) RDE configuration is presented to perform large-eddy simulations (LES) at 5 different inlet velocities, allowing for investigations into the performance of the three methods in predicting the dynamics changes as a result of parametric variations, as would be observed in real world applications. First, an \textit{a priori} analysis is performed to assess the capability of representative ROMs constructed using a linear static basis and nonlinear quadratic basis in modelling the RDE dynamics by performing a full order model (FOM) simulation. The two bases are constructed using training data generated from three of the five inlet velocity cases ($V_{in} = \text{100, 150, and 200 m/s}$). Evaluations of these methods reveal that the construction utilizing nonlinear quadratic basis exhibits improvements over linear static basis in low-rank approximation of the RDE dynamics over the training regime, but demonstrated restricted capabilites in accurately capturing the physics outside of the training dataset. This restricted capability is apparent in both future state and parametric prediction, indicating limited flexibility in constructing a truly predictive ROM for parametric evaluations of RDEs. Second, the newly developed adaptive MOR formulation is leveraged to construct an online ROM for 2D RDE modelling. The adaptive MOR is shown to require a small fraction of the training dataset for just one inlet velocity case, and demonstrates improved low-rank approximation when compared to the linear static basis method in training regime reconstruction, future state, and parametric prediction. This includes capturing the transient effects related to inlet velocity variations, a challenging characteristic for ROMs to capture. Additionally, a factor of 4 computational acceleration is shown when compared to the FOM simulation. Third, the ROM constructed from the adaptive MOR method is applied at every time near the initiation of the FOM simulation, where it is shown to accurately capture the transient RDE dynamics present prior to the establishment of a steady detonation wave. This capability allows for a significant reduction in the runtime of the FOM, allowing for further computational cost savings. The future work will focus on implementing proper dynamic load balancing based on the adapted sampling-point distributions to lift the current bottle neck in computational efficiency gain through the adaptive ROM. In addition, the scalability of the adaptive ROM method to large-scale 3D RDE applications will be another focus of the future work and an important stepping stone towards real-scale engineering applications.

\section{Acknowledgements}
This work was made possible with support from the Air Force Center of Excellence under grant FA9550-17-1-0195, titled "Multi-Fidelity Modeling of Rocket Combustor Dyanmics" (Technical Monitors: Fariba Fahroo, Mitat Birkan, Ramakanth Munipalli). Additional support was also provided by the SMART Scholarship Program through the sponsorship of the Naval Air Warfare Center - Weapons Division, China Lake facility. High performance computing capabilities were possible thanks to support from the Center for Research Computing at the University of Kansas.

\appendix
\section{ Governing Equations for Full Order Model}
\label{appendix:fom_eq}

The full order model computations are carried out with an in-house CFD code, the General Equations and Mesh Solver (GEMS), which solves the conservation equations for mass, momentum, energy and species mass fractions in a coupled fashion,
\begin{equation}
    \frac{\partial{Q}}{\partial{t}}+\nabla \cdot\left(\vec{F}-\vec{F_v}\right)=H,
    \label{eq:fom:governing}
\end{equation}
where $Q$ is the vector of conserved variables defined as, $Q=\left(\begin{array}{cccccc}
        \rho & \rho{u} & \rho{v} & \rho{w} & \rho{h^0-p} & \rho{Y_l}\\
    \end{array}\right)^T$
with $\rho$ representing density, $u$, $v$ and $w$ representing velocity field, $Y_l$ representing the $l^{th}$ species mass fraction and the total enthalpy $h^0$ is defined as, $h^0=h+\frac{1}{2}(u^2_i)=\sum_l{h_l{Y_l}}+\frac{1}{2}(u^2_i)$.

The fluxes have been separated into inviscid, $\vec{F}=F_{i}\vec{i}+F_{j}\vec{j}+F_{k}\vec{k}$ and viscous terms, $\vec{F_v}=F_{v,i}\vec{i}+F_{v,j}\vec{j}+F_{v,k}\vec{k}$. And the three inviscid fluxes are,
\begin{equation}
    F_i = \left(\begin{array}{c}
        \rho{u} \\
        \rho{u^2}+p \\
        \rho{uv} \\
        \rho{uw} \\
        \rho{uh^0} \\
        \rho{uY_l} \\
    \end{array} \right), \; F_j = \left(\begin{array}{c}
        \rho{v} \\
        \rho{uv} \\
        \rho{v^2}+p \\
        \rho{vw} \\
        \rho{vh^0} \\
        \rho{vY_l} \\
    \end{array} \right) \; \text{and} \; F_k = \left(\begin{array}{c}
        \rho{w} \\
        \rho{uw} \\
        \rho{vw}+p \\
        \rho{w^2}+p \\
        \rho{wh^0} \\
        \rho{wY_l} \\
    \end{array} \right)
    \label{eq:fom:inviscid_fluxes}
\end{equation}
The viscous fluxes are,
\begin{equation}
    \scalebox{0.8}{$F_{v,i} = \left(\begin{array}{c}
        0 \\
        \tau_{ii} \\
        \tau_{ji} \\
        \tau_{ki} \\
        u\tau_{ii}+v\tau_{ji}+w\tau_{ki}-q_i \\
        \rho{D_{l}}\frac{\partial{Y_l}}{\partial{x}} \\
    \end{array} \right), \; F_{v,j} = \left(\begin{array}{c}
        0 \\
        \tau_{ij} \\
        \tau_{jj} \\
        \tau_{kj} \\
        u\tau_{ij}+v\tau_{jj}+w\tau_{kj}-q_j \\
        \rho{D_{l}}\frac{\partial{Y_l}}{\partial{y}} \\
    \end{array} \right) \; \text{and} \; F_{v,k} = \left(\begin{array}{c}
        0 \\
        \tau_{ik} \\
        \tau_{jk} \\
        \tau_{kk} \\
        u\tau_{ik}+v\tau_{jk}+w\tau_{kk}-q_k \\
        \rho{D_{l}}\frac{\partial{Y_l}}{\partial{z}} \\
    \end{array} \right)$}
    \label{eq:fom:viscous_fluxes}
\end{equation}
where $D_l$ is defined to be the diffusion of the $l^{th}$  species into the mixture. In practice, this is an approximation used to model the multicomponent diffusion as the binary diffusion of each species into a mixture.

The heat flux in the $i^{th}$ direction, $q_i$, is defined as,
\begin{equation}
    q_i = -K\frac{\partial{T}}{\partial{x_i}}+\rho\sum^N_{l=1}D_l\frac{\partial{Y_l}}{\partial{x_i}}h_l+\mathbf{Q}_{\text{source}}
    \label{eq:fom:heat_flux}
\end{equation}
The three terms in the heat flux represent the heat transfer due to the conduction, species diffusion and heat generation from a volumetric source (e.g. heat radiation or external heat source) respectively.

The shear stress, $\tau$ , is also found in the viscous flux and defined in terms of the molecular viscosity and velocity field,
\begin{equation}
    \tau_{ij} = \mu\left(\frac{\partial{u_i}}{\partial{x_j}}+\frac{\partial{u_j}}{\partial{x_i}}-\frac{2}{3}\frac{\partial{u_m}}{\partial{x_m}}\delta_{ij}\right)
    \label{eq:fom:shear_stress}
\end{equation}

The source term, $H$ includes a single entry for each of the species equations signifying the production or destruction of the $l^{th}$ species, $\dot{\omega}_l$, which is determined by the chemical kinetics~\cite{WestbrookDryer},
\begin{equation}
    H=\left(\begin{array}{cccccc}
        0 & 0 & 0 & 0 & 0 & \dot{\omega}_l\\
    \end{array}\right)^T
    \label{eq:fom:source_term}
\end{equation}
\section{\textit{A priori} Analysis of Training Data Effects}
\label{appendix:training}

\begin{figure}[hbt!]
\centering
\includegraphics[width=.55\textwidth]{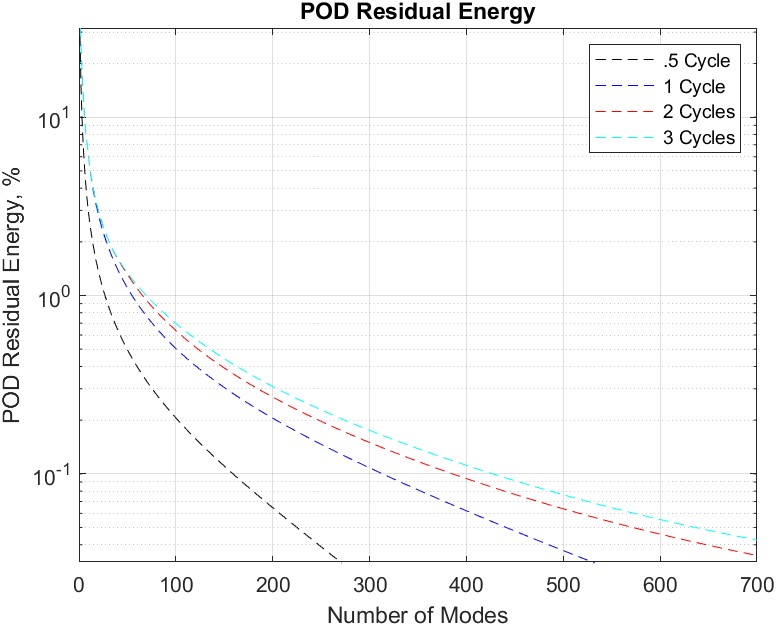}
\caption{Residual Energy Capture for Different Numbers of Simulation Cycles}\label{fig:pod_res}
\end{figure}

 Different number of FOM snapshots (beginning at 1ms) are included in the training dataset for the case where $V_{in} = 150\ m/s$ to compute a POD trial basis, corresponding to 0.5, 1, 2, and 3 cycles of snapshots. As seen in Figure ~\ref{fig:pod_res}, the POD residual energy exhibits slow decay, indicated by the increasingly large number of POD modes required to recover more of the total energy (0.5 cycle: 159 modes, 1 cycle: 316 modes, 2 cycles: 386 modes, and 3 cycles with 428 modes to recover 99.9\% of the energy of the system). As discussed in section~\ref{sec:intro}, this is a direct indicator of the slow decay of Kolmogorov N-width~\citep{Greif_2019}, which has been observed in convection-dominated multi-scale problems studied previously~\citep{Huang_2023,HuangMPLSVT2022,McQuarrieOpInf2021}. Unlike the common non-convergence characteristics of the POD residual energy observed in these previous studies, however, the residual energy here shows a trend of convergence when increasing the number of FOM snapshots in the training dataset. For example, the number of POD modes required to recover 99.9\% of the total energy increases less as more cycles of snapshots are included. Such convergence of the POD residual energy can be attributed to the dominance of detonation wave over other physics (e.g. turbulence) in the present test problem in Fig.~\ref{fig:FOM_Results}. More importantly, it indicates that it is feasible to represent the overall (e.g. 5 cycles) detonation wave dynamics by a POD basis computed using the smaller number of training snapshots.

We quantify the capability of the generated POD basis in representing the target dynamics via the projected FOM solution, $\Bar{\mathbf{q}}_p$, defined as follows for a conventional linear static basis
\begin{equation}
    \Bar{\mathbf{q}}_p = \mathbf{q}_{p,\text{ref}} + \scaleMatPrim^{-1} \trialBasisPrim \trialBasisPrim^T \mathbf{q}_p .
    \label{eq:projFOM}
\end{equation}
where $\mathbf{q}_p$ represents the state solution obtained from the FOM. It is noteworthy that the projected FOM solution provides a \textit{best-scenario} evaluation of the accuracy of the resulting ROM (i.e. the ROM solution cannot produce more accurate predictions of the dynamics than the projected FOM solution)~\citep{Arnold-Medalbalimi_2022}.

\begin{figure}[hbt!]
    \centering
    \begin{subfigure}[b]{0.45\textwidth}
         \centering
         \includegraphics[width=\textwidth]{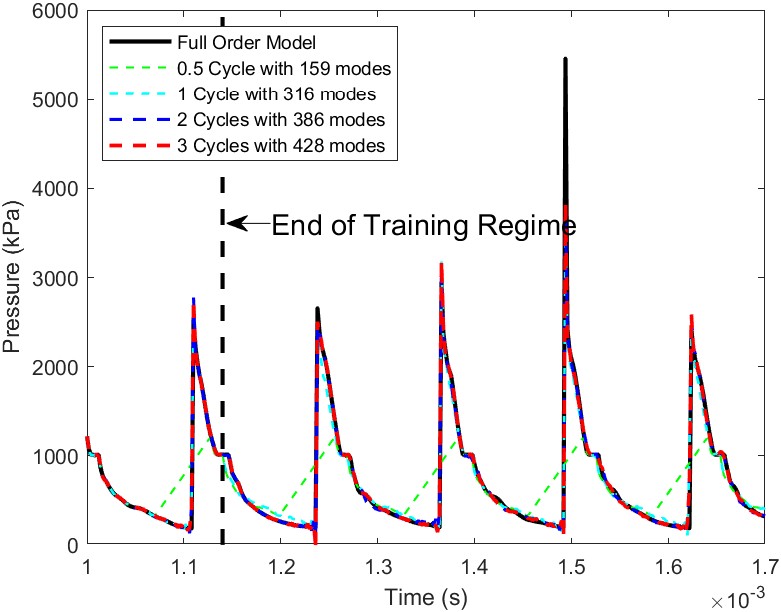}
         \caption{99.9\% Residual Energy Capture}
         \label{fig:PO_fig}
     \end{subfigure}
    \begin{subfigure}[b]{0.45\textwidth}
         \centering
         \includegraphics[width=\textwidth]{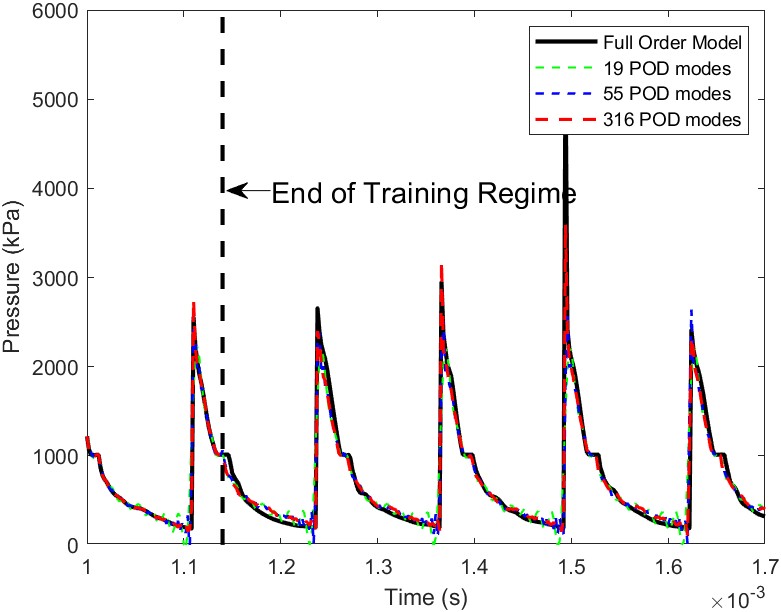}
         \caption{1 Cycle of Training Data}
         \label{fig:PO1Run_fig}
     \end{subfigure}
     \hfill
        \caption{Comparisons of local pressure trace between FOM and the projected FOM for $\mathbf{V_{in}}$  = 150 m/s.}
        \label{fig:projectedFOMFuture}
\end{figure}

In order to determine the viability of traditional PROMs for future state prediction, we first investigate the effects on projected solution accuracy for linear static bases generated with different training data sets. The results in Fig.~\ref{fig:projectedFOMFuture} show graphically the time evolution of pressure over the cell located at the bottom-most left-hand side of domain for the FOM and projected FOM solutions. Fig.~\ref{fig:PO_fig} focuses on evaluating the representativeness of the basis in the future-state prediction with mode numbers corresponding to 99.9\% POD energy in Fig.~\ref{fig:pod_res}. Here the importance of ensuring that adequate offline training data is used ($\ge$ 1 cycle) is demonstrated. We see that if less than this amount of training data is used (i.e., 0.5 cycle), the projected FOM solution does not provide useful results outside of the space traversed in the training regime. Additionally, while increasing the number of cycles included in the training data does improve accuracy, the improvements for training sets larger than 1 cycle are shown to be marginal, with the benefits arguably outweighed by the added offline computational and storage costs. This indicates, therefore, that the most economical training set is 1 cycle (7000 time snapshots). 

Having shown that a single cycle of training data is sufficient, it is now useful to determine to what degree the amount of included modes affects the accuracy of the solution. For mode numbers corresponding to 97\%, 99\%, and 99.9\% residual energy capture, the resulting projected FOM solutions are shown in Fig.~\ref{fig:PO1Run_fig}. As expected, the more POD modes are included, the more accurately the FOM solution is recreated, with 316 modes corresponding to 99.9\% energy capture. Although the energy capture is only increasing by 2.9\% across the three bases, the required modes increase from 19 to 316, again illustrating the slow Kolmogorov N-width decay exhibited by this problem. 

\begin{figure}[hbt!]
    \centering
    \begin{subfigure}[b]{0.55\textwidth}
         \centering
         \includegraphics[width=\textwidth]{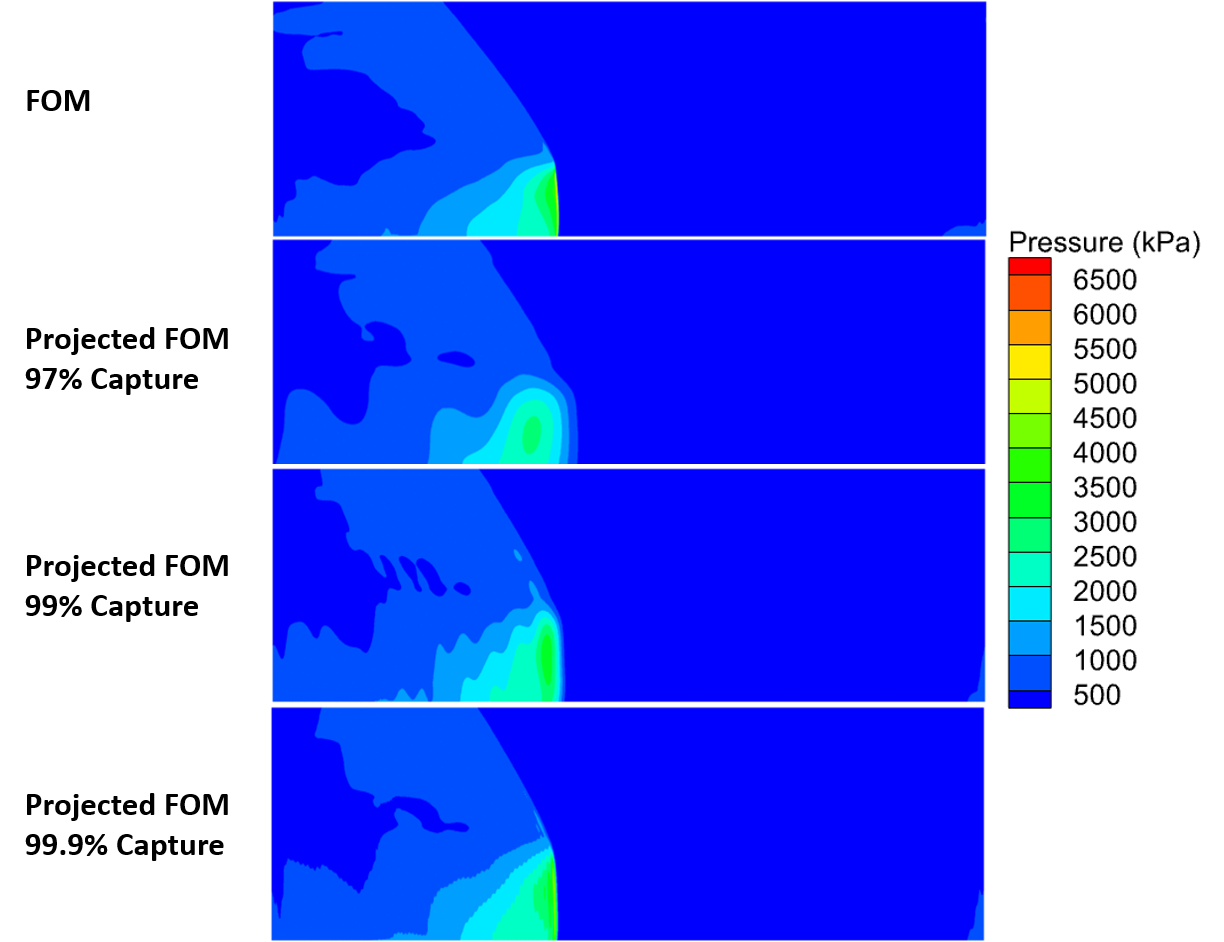}
     \end{subfigure}
     \hfill
        \caption{Comparisons of pressure field between FOM and projected FOM at $t  = 1.16 \ \text{ms}$.}
        \label{fig:PTrain}
\end{figure}

Although Fig.~\ref{fig:PO1Run_fig} indicates the most accurate solution is provided by the linear basis with 316 modes as expected, all three test cases capture the important structures of the detonation dynamics. Therefore, in order to more rigorously determine the true number of modes required for this problem, the flow field for both the FOM and projected FOM was generated, and can be seen in Fig. ~\ref{fig:PTrain}. The snapshots shown were taken outside of the training regime ($t = 1.16 ms$) and reveal that it is indeed necessary to utilize modes corresponding to at least 99.9 \% residual energy capture.

\section{Regularization Parameter}
\label{appendix:regularization}

\begin{figure}
    \centering
    \includegraphics[width=.5\textwidth]{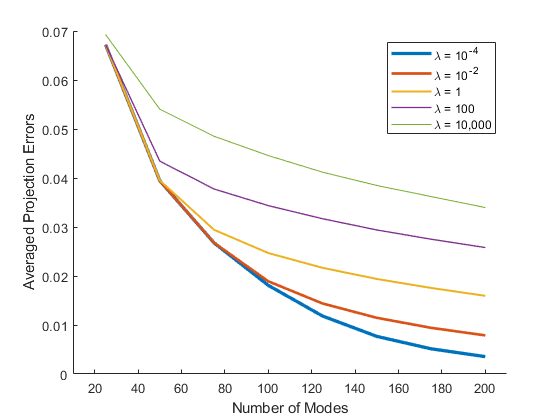}
    \caption{Comparisons of the projection errors using quadratic bases with different regularization parameters, ($\lambda$), in reproducing the training dataset.}
    \label{fig:NLErrorComparison}
\end{figure}

Special consideration needs to be taken when determining a proper scalar regularization parameter, ($\lambda$), as this term is responsible for inhibiting overfitting of the system operators to the data \citep{McQuarrieOpInf2021,Geelen_2023_Quadratic}. This, in turn, promotes stability and accuracy of the solutions generated with the nonlinear quadratic method. In order to determine a proper value for $\lambda$, the typical procedure involves testing a range of values when generating the nonlinear quadratic basis to observe how the resulting basis is affected. In the case of this study, a range of values from $10^{-4}$ to $10^4$ are used to generate different nonlinear quadratic bases. The resulting projection errors are averaged over the three training cases using \ref{eq:pod:proj_err} and compared in Figure~\ref{fig:NLErrorComparison}. It can readily be seen that the accuracy of the quadratic basis in this case is sensitive to the selection of the regularization parameter, and as $\lambda$ increases, the projection errors of the quadratic basis increase as well. It can also be seen that as $\lambda \rightarrow 0$, the projection errors begin to converge, indicating that lower values of $\lambda$ will provide diminished improvements in accuracy. Due to the imposed cap on the number of included linear modes being 200, and the observed trend of convergence, it was determined that $\lambda = 10^{-4}$ would be suitable for this study. 

\clearpage
\bibliography{sample}

\begin{thebibliography}{76}
\newcommand{\enquote}[1]{``#1''}
\providecommand{\natexlab}[1]{#1}
\providecommand{\url}[1]{\texttt{#1}}
\providecommand{\urlprefix}{URL }
\expandafter\ifx\csname urlstyle\endcsname\relax
  \providecommand{\doi}[1]{\discretionary{}{}{}https://doi.org/#1}\else
  \providecommand{\doi}[1]{\discretionary{}{}{}\urlstyle{rm}\url{https://doi.org/#1}}\fi

\bibitem[{Adamson and Olsson(1967)}]{Adamson_1967}
Adamson, T., and Olsson, G., \enquote{Performance analysis of rotating detonation wave rocket engine,} \emph{Acta Astronautica}, Vol.~13, 1967, pp. 405--415.

\bibitem[{Zeldovich(2006)}]{Zeldovich_1940}
Zeldovich, Y.~B., \enquote{To the Question of Energy Use of Detonation Combustion,} \emph{Journal of Propulsion and Power}, Vol.~22, No.~3, 2006, pp. 588--592.
\newblock \doi{10.2514/1.22705}.

\bibitem[{Anand et~al.(2016)Anand, St.~George, Driscoll, and Gutmark}]{Anand_2016}
Anand, V., St.~George, A., Driscoll, R., and Gutmark, E., \enquote{Investigation of rotating detonation combustor operation with H 2 -Air mixtures,} \emph{International Journal of Hydrogen Energy}, Vol.~41, No.~2, 2016, pp. 1281--1292.
\newblock \doi{10.1016/j.ijhydene.2015.11.041}.

\bibitem[{Fotia et~al.(2016)Fotia, Schauer, Kaemming, and Hoke}]{Fotia_2016}
Fotia, M.~L., Schauer, F., Kaemming, T., and Hoke, J., \enquote{Experimental Study of the Performance of a Rotating Detonation Engine with Nozzle,} \emph{Journal of Propulsion and Power}, Vol.~32, No.~3, 2016, pp. 674--681.
\newblock \doi{10.2514/1.B35913}.

\bibitem[{Nordeen et~al.(2014)Nordeen, Schwer, Schauer, Hoke, Barber, and Cetegen}]{Nordeen_2014}
Nordeen, C.~A., Schwer, D., Schauer, F., Hoke, J., Barber, T., and Cetegen, B., \enquote{Thermodynamic model of a rotating detonation engine,} \emph{Combustion, Explosion, and Shock Waves}, Vol.~50, No.~5, 2014, pp. 568--577.
\newblock \doi{10.1134/s0010508214050128}.

\bibitem[{Shao et~al.(2010)Shao, Liu, and Wang}]{Shao_2010}
Shao, Y.-T., Liu, M., and Wang, J.-P., \enquote{Numerical Investigation of Rotating Detonation Engine Propulsive Performance,} \emph{Combustion Science and Technology}, Vol. 182, No. 11-12, 2010, pp. 1586--1597.
\newblock \doi{10.1080/00102202.2010.497316}.

\bibitem[{Sousa et~al.(2017)Sousa, Paniagua, and Collado~Morata}]{Sousa_2017}
Sousa, J., Paniagua, G., and Collado~Morata, E., \enquote{Thermodynamic analysis of a gas turbine engine with a rotating detonation combustor,} \emph{Applied Energy}, Vol. 195, 2017, pp. 247--256.
\newblock \doi{10.1016/j.apenergy.2017.03.045}.

\bibitem[{Rankin et~al.(2017)Rankin, Fotia, Naples, Stevens, Hoke, Kaemming, Theuerkauf, and Schauer}]{Rankin_2017}
Rankin, B.~A., Fotia, M.~L., Naples, A.~G., Stevens, C.~A., Hoke, J.~L., Kaemming, T.~A., Theuerkauf, S.~W., and Schauer, F.~R., \enquote{Overview of Performance, Application, and Analysis of Rotating Detonation Engine Technologies,} \emph{Journal of Propulsion and Power}, Vol.~33, No.~1, 2017, pp. 131--143.
\newblock \doi{10.2514/1.B36303}.

\bibitem[{Bennewitz et~al.(2019)Bennewitz, Bigler, Pilgram, and Hargus}]{Bennewitz_2019}
Bennewitz, J.~W., Bigler, B.~R., Pilgram, J.~J., and Hargus, W.~A., \enquote{Modal Transitions in Rotating Detonation Rocket Engines,} \emph{International Journal of Energetic Materials and Chemical Propulsion}, Vol.~18, 2019, pp. 91--109.

\bibitem[{Kindracki et~al.(2011)Kindracki, Wolański, and Gut}]{Kindracki_2011}
Kindracki, J., Wolański, P., and Gut, Z., \enquote{Experimental research on the rotating detonation in gaseous fuels–oxygen mixtures,} \emph{Shock Waves}, Vol.~21, No.~2, 2011, pp. 75--84.
\newblock \doi{10.1007/s00193-011-0298-y}.

\bibitem[{Walters et~al.(2021)Walters, Gejji, Heister, and Slabaugh}]{Walters_2021}
Walters, I.~V., Gejji, R.~M., Heister, S.~D., and Slabaugh, C.~D., \enquote{Flow and performance analysis of a natural gas-air rotating detonation engine with high-speed velocimetry,} \emph{Combustion and Flame}, Vol. 232, 2021.
\newblock \doi{10.1016/j.combustflame.2021.111549}.

\bibitem[{Pal et~al.(2021)Pal, Demir, Kundu, and Som}]{Pal_LESRDE}
Pal, P., Demir, S., Kundu, P., and Som, S., \enquote{Large-Eddy Simulations of Methane-Oxygen Combustion in a Rotating Detonation Rocket Engine,} , 2021.
\newblock \doi{10.2514/6.2021-3642}.

\bibitem[{Prakash et~al.(2021)Prakash, Raman, Lietz, Hargus, and Schumaker}]{Prakash_2021}
Prakash, S., Raman, V., Lietz, C.~F., Hargus, W.~A., and Schumaker, S.~A., \enquote{Numerical simulation of a methane-oxygen rotating detonation rocket engine,} \emph{Proceedings of the Combustion Institute}, Vol.~38, No.~3, 2021, pp. 3777--3786.
\newblock \doi{10.1016/j.proci.2020.06.288}.

\bibitem[{Lietz et~al.(2020)Lietz, Ross, Desai, and Hargus}]{Lietz_2020}
Lietz, C., Ross, M., Desai, Y., and Hargus, W.~A., \enquote{Numerical investigation of operational performance in a methane-oxygen rotating detonation rocket engine,} , 2020.
\newblock \doi{10.2514/6.2020-0687}.

\bibitem[{Schwer et~al.(2019)Schwer, Johnson, Kercher, Kessler, and Corrigan}]{Schwer_2019}
Schwer, D.~A., Johnson, R.~F., Kercher, A., Kessler, D., and Corrigan, A.~T., \enquote{Progress in Efficient, High-Fidelity, Rotating Detonation Engine Simulations,} , 2019.
\newblock \doi{10.2514/6.2019-2018}.

\bibitem[{Sato et~al.(2021)Sato, Chacon, Gamba, and Raman}]{Sato_2021_MassFlow}
Sato, T., Chacon, F., Gamba, M., and Raman, V., \enquote{Mass flow rate effect on a rotating detonation combustor with an axial air injection,} \emph{Shock Waves}, Vol.~31, No.~7, 2021, pp. 741--751.
\newblock \doi{10.1007/s00193-020-00984-7}.

\bibitem[{Rood et~al.(2021)Rood, Henry~de Frahan, Day, Sitaraman, Yellapantula, Perry, Grout, Almgren, Zhang, and Chen}]{Rood_et_al}
Rood, J.~S., Henry~de Frahan, M.~T., Day, M.~S., Sitaraman, H., Yellapantula, S., Perry, B.~A., Grout, R.~W., Almgren, A., Zhang, W., and Chen, J., \enquote{Enabling Combustion Science Simulations for Future Exascale Machines,} 2021.
\newblock \urlprefix\url{https://www.osti.gov/biblio/1833338}.

\bibitem[{Lietz et~al.(2019)Lietz, Desai, Munipalli, Schumaker, and Sankaran}]{Lietz_2019}
Lietz, C., Desai, Y., Munipalli, R., Schumaker, S.~A., and Sankaran, V., \enquote{Flowfield analysis of a 3D simulation of a rotating detonation rocket engine,} , 2019.
\newblock \doi{10.2514/6.2019-1009}.

\bibitem[{Kaemming et~al.(2017)Kaemming, Fotia, Hoke, and Schauer}]{Kaemming_2017}
Kaemming, T., Fotia, M.~L., Hoke, J., and Schauer, F., \enquote{Thermodynamic Modeling of a Rotating Detonation Engine Through a Reduced-Order Approach,} \emph{Journal of Propulsion and Power}, Vol.~33, No.~5, 2017, pp. 1170--1178.
\newblock \doi{10.2514/1.B36237}.

\bibitem[{Koch(2021)}]{Koch_2021}
Koch, J., \enquote{Data-driven surrogates of rotating detonation engine physics with neural ordinary differential equations and high-speed camera footage,} \emph{Physics of Fluids}, Vol.~33, No.~9, 2021.
\newblock \doi{10.1063/5.0063624}.

\bibitem[{Zhao et~al.(2021)Zhao, Cleary, and Zhang}]{Zhao_2021}
Zhao, M., Cleary, M.~J., and Zhang, H., \enquote{Combustion mode and wave multiplicity in rotating detonative combustion with separate reactant injection,} \emph{Combustion and Flame}, Vol. 225, 2021, pp. 291--304.
\newblock \doi{10.1016/j.combustflame.2020.11.001}.

\bibitem[{Mahmoudi et~al.(2014)Mahmoudi, Karimi, Deiterding, and Emami}]{Mahmoudi_2014}
Mahmoudi, Y., Karimi, N., Deiterding, R., and Emami, S., \enquote{Hydrodynamic Instabilities in Gaseous Detonations: Comparison of Euler, Navier–Stokes, and Large-Eddy Simulation,} \emph{Journal of Propulsion and Power}, Vol.~30, No.~2, 2014, pp. 384--396.
\newblock \doi{10.2514/1.B34986}.

\bibitem[{Koch et~al.(2020)Koch, Kurosaka, Knowlen, and Kutz}]{Koch_2020ModeLocked}
Koch, J., Kurosaka, M., Knowlen, C., and Kutz, J.~N., \enquote{Mode-locked rotating detonation waves: Experiments and a model equation,} \emph{Phys Rev E}, Vol. 101, No. 1-1, 2020, p. 013106.
\newblock \doi{10.1103/PhysRevE.101.013106}, \urlprefix\url{https://www.ncbi.nlm.nih.gov/pubmed/32069601}.

\bibitem[{Koch et~al.(2021)Koch, Kurosaka, Knowlen, and Kutz}]{Koch_2021RDE}
Koch, J., Kurosaka, M., Knowlen, C., and Kutz, J.~N., \enquote{Multiscale physics of rotating detonation waves: Autosolitons and modulational instabilities,} \emph{Phys Rev E}, Vol. 104, No. 2-1, 2021, p. 024210.
\newblock \doi{10.1103/PhysRevE.104.024210}, \urlprefix\url{https://www.ncbi.nlm.nih.gov/pubmed/34525544}.

\bibitem[{Cherkassky and Mulier(2007)}]{cherkassky2007learning}
Cherkassky, V., and Mulier, F., \emph{Learning from Data: Concepts, Theory, and Methods}, IEEE Press, Wiley, 2007.
\newblock \urlprefix\url{https://books.google.com/books?id=IMGzP-IIaKAC}.

\bibitem[{Koch and Kutz(2020)}]{Koch_2020}
Koch, J., and Kutz, J.~N., \enquote{Modeling thermodynamic trends of rotating detonation engines,} \emph{Physics of Fluids}, Vol.~32, No.~12, 2020.
\newblock \doi{10.1063/5.0023972}.

\bibitem[{Mendible et~al.(2021)Mendible, Koch, Lange, Brunton, and Kutz}]{Mendible_2021}
Mendible, A., Koch, J., Lange, H., Brunton, S.~L., and Kutz, J.~N., \enquote{Data-driven modeling of rotating detonation waves,} \emph{Physical Review Fluids}, Vol.~6, No.~5, 2021.
\newblock \doi{10.1103/PhysRevFluids.6.050507}.

\bibitem[{Zhou et~al.(2022)Zhou, Song, Ji, and Wei}]{Zhou_2022}
Zhou, L., Song, Y., Ji, W., and Wei, H., \enquote{Machine learning for combustion,} \emph{Energy and AI}, Vol.~7, 2022.
\newblock \doi{10.1016/j.egyai.2021.100128}.

\bibitem[{Duraisamy(2021)}]{Duraisamy_2021}
Duraisamy, K., \enquote{Perspectives on machine learning-augmented Reynolds-averaged and large eddy simulation models of turbulence,} \emph{Physical Review Fluids}, Vol.~6, No.~5, 2021.
\newblock \doi{10.1103/PhysRevFluids.6.050504}.

\bibitem[{Ihme et~al.(2022)Ihme, Chung, and Mishra}]{Ihme_2022}
Ihme, M., Chung, W.~T., and Mishra, A.~A., \enquote{Combustion machine learning: Principles, progress and prospects,} \emph{Progress in Energy and Combustion Science}, Vol.~91, 2022.
\newblock \doi{10.1016/j.pecs.2022.101010}.

\bibitem[{Raissi et~al.(2019)Raissi, Perdikaris, and Karniadakis}]{Raissi_2019}
Raissi, M., Perdikaris, P., and Karniadakis, G.~E., \enquote{Physics-informed neural networks: A deep learning framework for solving forward and inverse problems involving nonlinear partial differential equations,} \emph{Journal of Computational Physics}, Vol. 378, 2019, pp. 686--707.
\newblock \doi{10.1016/j.jcp.2018.10.045}.

\bibitem[{Benner et~al.(2015)Benner, Gugercin, and Willcox}]{Benner_2015}
Benner, P., Gugercin, S., and Willcox, K., \enquote{A Survey of Projection-Based Model Reduction Methods for Parametric Dynamical Systems,} \emph{SIAM Review}, Vol.~57, No.~4, 2015, pp. 483--531.
\newblock \doi{10.1137/130932715}.

\bibitem[{Peherstorfer and Willcox(2016)}]{Peherstorfer_2016}
Peherstorfer, B., and Willcox, K., \enquote{Data-driven operator inference for nonintrusive projection-based model reduction,} \emph{Computer Methods in Applied Mechanics and Engineering}, Vol. 306, 2016, pp. 196--215.
\newblock \doi{10.1016/j.cma.2016.03.025}.

\bibitem[{Kramer and Willcox(2019)}]{Kramer_2019}
Kramer, B., and Willcox, K.~E., \enquote{Nonlinear Model Order Reduction via Lifting Transformations and Proper Orthogonal Decomposition,} \emph{AIAA Journal}, Vol.~57, No.~6, 2019, pp. 2297--2307.
\newblock \doi{10.2514/1.J057791}.

\bibitem[{McQuarrie et~al.(2021)McQuarrie, Huang, and Willcox}]{McQuarrieOpInf2021}
McQuarrie, S.~A., Huang, C., and Willcox, K.~E., \enquote{Data-driven reduced-order models via regularised Operator Inference for a single-injector combustion process,} \emph{Journal of the Royal Society of New Zealand}, Vol.~51, No.~2, 2021, pp. 194--211.
\newblock \doi{10.1080/03036758.2020.1863237}.

\bibitem[{Barbagallo et~al.(2012)Barbagallo, Dergham, Sipp, Schmid, and Robinet}]{Barbagallo_2012}
Barbagallo, A., Dergham, G., Sipp, D., Schmid, P.~J., and Robinet, J.-C., \enquote{Closed-loop control of unsteadiness over a rounded backward-facing step,} \emph{Journal of Fluid Mechanics}, Vol. 703, 2012, pp. 326--362.
\newblock \doi{10.1017/jfm.2012.223}.

\bibitem[{Lieu and Farhat(2007)}]{Lieu_2007}
Lieu, T., and Farhat, C., \enquote{Adaptation of Aeroelastic Reduced-Order Models and Application to an F-16 Configuration,} \emph{AIAA Journal}, Vol.~45, No.~6, 2007, pp. 1244--1257.
\newblock \doi{10.2514/1.24512}.

\bibitem[{Blonigan et~al.(2021)Blonigan, Rizzi, Howard, Fike, and Carlberg}]{Blonigan_2021}
Blonigan, P.~J., Rizzi, F., Howard, M., Fike, J.~A., and Carlberg, K.~T., \enquote{Model Reduction for Steady Hypersonic Aerodynamics via Conservative Manifold Least-Squares Petrov–Galerkin Projection,} \emph{AIAA Journal}, Vol.~59, No.~4, 2021, pp. 1296--1312.
\newblock \doi{10.2514/1.J059785}.

\bibitem[{Huang et~al.(2022{\natexlab{a}})Huang, Wentland, Duraisamy, and Merkle}]{HuangMPLSVT2022}
Huang, C., Wentland, C.~R., Duraisamy, K., and Merkle, C., \enquote{Model reduction for multi-scale transport problems using model-form preserving least-squares projections with variable transformation,} \emph{Journal of Computational Physics}, Vol. 448, 2022{\natexlab{a}}.
\newblock \doi{10.1016/j.jcp.2021.110742}.

\bibitem[{Farcas et~al.(2022)Farcas, Munipalli, and Willcox}]{Farcas_2022}
Farcas, I., Munipalli, R., and Willcox, K.~E., \enquote{On filtering in non-intrusive data-driven reduced-order modeling,} , 2022.
\newblock \doi{10.2514/6.2022-3487}.

\bibitem[{Farcas et~al.(2023)Farcas, Gundevia, Munipalli, and Willcox}]{farcas2023improving}
Farcas, I.-G., Gundevia, R.~P., Munipalli, R., and Willcox, K.~E., \enquote{Improving the accuracy and scalability of large-scale physics-based data-driven reduced modeling via domain decomposition,} , 2023.

\bibitem[{Lumley and Poje(1997)}]{Lumley1997}
Lumley, J.~L., and Poje, A., \enquote{Low-dimensional models for flows with density fluctuations,} \emph{Physics of Fluids}, Vol.~9, No.~7, 1997, p. 2023.
\newblock \doi{10.1063/1.869321}.

\bibitem[{Greif and Urban(2019)}]{Greif_2019}
Greif, C., and Urban, K., \enquote{Decay of the Kolmogorov N-width for wave problems,} \emph{Applied Mathematics Letters}, Vol.~96, 2019, pp. 216--222.
\newblock \doi{10.1016/j.aml.2019.05.013}.

\bibitem[{Geelen and Willcox(2022)}]{Geelen_2022}
Geelen, R., and Willcox, K., \enquote{Localized non-intrusive reduced-order modelling in the operator inference framework,} \emph{Philosophical Transactions of the Royal Society A: Mathematical, Physical and Engineering Sciences}, Vol. 380, 2022.
\newblock \doi{10.1098/rsta.2021.0206}.

\bibitem[{Amsallem et~al.(2015)Amsallem, Zahr, and Washabaugh}]{Amsallem_2015}
Amsallem, D., Zahr, M.~J., and Washabaugh, K., \enquote{Fast local reduced basis updates for the efficient reduction of nonlinear systems with hyper-reduction,} \emph{Advances in Computational Mathematics}, Vol.~41, No.~5, 2015, pp. 1187--1230.
\newblock \doi{10.1007/s10444-015-9409-0}.

\bibitem[{Ohlberger and Rave(2013)}]{Ohlberger_2013}
Ohlberger, M., and Rave, S., \enquote{Nonlinear reduced basis approximation of parameterized evolution equations via the method of freezing,} \emph{Comptes Rendus. Mathématique}, Vol. 351, No. 23-24, 2013, pp. 901--906.
\newblock \doi{10.1016/j.crma.2013.10.028}.

\bibitem[{Reiss et~al.(2018)Reiss, Schulze, Sesterhenn, and Mehrmann}]{Reiss_2018}
Reiss, J., Schulze, P., Sesterhenn, J., and Mehrmann, V., \enquote{The Shifted Proper Orthogonal Decomposition: A Mode Decomposition for Multiple Transport Phenomena,} \emph{SIAM Journal on Scientific Computing}, Vol.~40, No.~3, 2018, pp. A1322--A1344.
\newblock \doi{10.1137/17m1140571}.

\bibitem[{Schulze et~al.(2018)Schulze, Reiss, and Mehrmann}]{schulze2018model}
Schulze, P., Reiss, J., and Mehrmann, V., \enquote{Model Reduction for a Pulsed Detonation Combuster via Shifted Proper Orthogonal Decomposition,} , 2018.

\bibitem[{Reiss(2021)}]{Reiss_2021}
Reiss, J., \enquote{Optimization-Based Modal Decomposition for Systems with Multiple Transports,} \emph{SIAM Journal on Scientific Computing}, Vol.~43, No.~3, 2021, pp. A2079--A2101.
\newblock \doi{10.1137/20m1322005}.

\bibitem[{Rim et~al.(2023)Rim, Peherstorfer, and Mandli}]{Rim_2023}
Rim, D., Peherstorfer, B., and Mandli, K.~T., \enquote{Manifold Approximations via Transported Subspaces: Model Reduction for Transport-Dominated Problems,} \emph{SIAM Journal on Scientific Computing}, Vol.~45, No.~1, 2023, pp. A170--A199.
\newblock \doi{10.1137/20m1316998}.

\bibitem[{Nair and Balajewicz(2019)}]{Nair_2019}
Nair, N.~J., and Balajewicz, M., \enquote{Transported snapshot model order reduction approach for parametric, steady-state fluid flows containing parameter-dependent shocks,} \emph{International Journal for Numerical Methods in Engineering}, Vol. 117, No.~12, 2019, pp. 1234--1262.
\newblock \doi{https://doi.org/10.1002/nme.5998}, \urlprefix\url{https://onlinelibrary.wiley.com/doi/abs/10.1002/nme.5998}.

\bibitem[{Alireza~Mirhoseini and Zahr(2023)}]{Mirhoseini_2023}
Alireza~Mirhoseini, M., and Zahr, M.~J., \enquote{Model reduction of convection-dominated partial differential equations via optimization-based implicit feature tracking,} \emph{Journal of Computational Physics}, Vol. 473, 2023.
\newblock \doi{10.1016/j.jcp.2022.111739}.

\bibitem[{Kim et~al.(2022)Kim, Choi, Widemann, and Zohdi}]{KimChoiNonlinearMainfold}
Kim, Y., Choi, Y., Widemann, D., and Zohdi, T., \enquote{A fast and accurate physics-informed neural network reduced order model with shallow masked autoencoder,} \emph{Journal of Computational Physics}, Vol. 451, 2022.
\newblock \doi{10.1016/j.jcp.2021.110841}.

\bibitem[{Lee and Carlberg(2020)}]{LeeCarlbergNonlinearManifold}
Lee, K., and Carlberg, K.~T., \enquote{Model reduction of dynamical systems on nonlinear manifolds using deep convolutional autoencoders,} \emph{Journal of Computational Physics}, Vol. 404, 2020.
\newblock \doi{10.1016/j.jcp.2019.108973}.

\bibitem[{Geelen et~al.(2023)Geelen, Wright, and Willcox}]{Geelen_2023_Quadratic}
Geelen, R., Wright, S., and Willcox, K., \enquote{Operator inference for non-intrusive model reduction with quadratic manifolds,} \emph{Computer Methods in Applied Mechanics and Engineering}, Vol. 403, 2023.
\newblock \doi{10.1016/j.cma.2022.115717}.

\bibitem[{Barnett and Farhat(2022)}]{Barnett_QuadraticPROM2022}
Barnett, J., and Farhat, C., \enquote{Quadratic approximation manifold for mitigating the Kolmogorov barrier in nonlinear projection-based model order reduction,} \emph{Journal of Computational Physics}, Vol. 464, 2022.
\newblock \doi{10.1016/j.jcp.2022.111348}.

\bibitem[{Peherstorfer(2022)}]{Peherstorfer_2022_Kolmogorov}
Peherstorfer, B., \enquote{Breaking the Kolmogorov Barrier with Nonlinear Model Reduction,} \emph{Notices of the American Mathematical Society}, Vol.~69, No.~05, 2022.
\newblock \doi{10.1090/noti2475}.

\bibitem[{Zucatti and Zahr(2024)}]{Zucatti_2024}
Zucatti, V., and Zahr, M.~J., \enquote{An adaptive, training-free reduced-order model for convection-dominated problems based on hybrid snapshots,} \emph{International Journal for Numerical Methods in Fluids}, Vol.~96, No.~2, 2024, pp. 189--208.
\newblock \doi{https://doi.org/10.1002/fld.5240}, \urlprefix\url{https://onlinelibrary.wiley.com/doi/abs/10.1002/fld.5240}.

\bibitem[{Peherstorfer(2020)}]{PeherstorferADEIM}
Peherstorfer, B., \enquote{{Model reduction for transport-dominated problems via online adaptive bases and adaptive sampling},} \emph{SIAM Journal on Scientific Computing}, Vol.~42, No.~5, 2020, pp. A2803--A2836.
\newblock \doi{10.1137/19M1257275}.

\bibitem[{Ramezanian et~al.(2021)Ramezanian, Nouri, and Babaee}]{Ramezanian_2021}
Ramezanian, D., Nouri, A.~G., and Babaee, H., \enquote{On-the-fly reduced order modeling of passive and reactive species via time-dependent manifolds,} \emph{Computer Methods in Applied Mechanics and Engineering}, Vol. 382, 2021.
\newblock \doi{10.1016/j.cma.2021.113882}.

\bibitem[{Huang and Duraisamy(2023)}]{Huang_2023}
Huang, C., and Duraisamy, K., \enquote{Predictive reduced order modeling of chaotic multi-scale problems using adaptively sampled projections,} \emph{Journal of Computational Physics}, Vol. 491, 2023.
\newblock \doi{10.1016/j.jcp.2023.112356}.

\bibitem[{Butcher(2016)}]{ButcherNumMeth}
Butcher, J., \emph{Numerical Methods for Ordinary Differential Equations}, John Wiley \& Sons, Ltd, 2016, Chap.~4, pp. 333--387.
\newblock \doi{https://doi.org/10.1002/9781119121534.ch4}, \urlprefix\url{https://onlinelibrary.wiley.com/doi/abs/10.1002/9781119121534.ch4}.

\bibitem[{Lumley(1967)}]{lumley1967structure}
Lumley, J.~L., \enquote{The structure of inhomogeneous turbulent flows,} \emph{Atmospheric turbulence and radio wave propagation}, 1967, pp. 166--178.

\bibitem[{Berkooz et~al.(1993)Berkooz, Holmes, and Lumley}]{Berkooz_1993}
Berkooz, G., Holmes, P., and Lumley, J.~L., \enquote{The proper orthogonal decomposition in the analysis of turbulent flows,} \emph{Annu. Rev. Fluid Mech}, Vol.~25, 1993, pp. 539--75.

\bibitem[{Everson and Sirovich(1995)}]{Everson1995_gappyPOD}
Everson, R., and Sirovich, L., \enquote{The {K}arhunen--{L}oeve procedure for gappy data,} \emph{Journal of the Optical Society of America A}, Vol.~12, 1995.
\newblock \doi{10.1364/JOSAA.12.001657}.

\bibitem[{Zimmermann et~al.(2018)Zimmermann, Peherstorfer, and Willcox}]{ZimmermannAdaptiveBasis2018}
Zimmermann, R., Peherstorfer, B., and Willcox, K., \enquote{Geometric Subspace Updates with Applications to Online Adaptive Nonlinear Model Reduction,} \emph{SIAM J. MATRIX ANAL. APPL.}, Vol.~39, 2018, pp. 234--261.

\bibitem[{Huang et~al.(2022{\natexlab{b}})Huang, Duraisamy, and Merkle}]{Huang_CBROM2022}
Huang, C., Duraisamy, K., and Merkle, C., \enquote{Component-based Reduced Order Modeling of Large-scale Complex Systems,} \emph{Frontiers in Physics}, 2022{\natexlab{b}}.

\bibitem[{Schwer et~al.(2014)Schwer, Corrigan, and Kailasanath}]{Schwer_2014}
Schwer, D.~A., Corrigan, A.~T., and Kailasanath, K., \enquote{Towards Efficient, Unsteady, Three-Dimensional Rotating Detonation Engine Simulations,} , 2014.
\newblock \doi{10.2514/6.2014-1014}.

\bibitem[{Schwer and Kailasanath(2011)}]{Schwer_2011}
Schwer, D., and Kailasanath, K., \enquote{Numerical investigation of the physics of rotating-detonation-engines,} \emph{Proceedings of the Combustion Institute}, Vol.~33, No.~2, 2011, pp. 2195--2202.
\newblock \doi{10.1016/j.proci.2010.07.050}.

\bibitem[{Harvazinski et~al.(2015)Harvazinski, Huang, Sankaran, Feldman, Anderson, Merkle, and Talley}]{Harvazinski_2015}
Harvazinski, M.~E., Huang, C., Sankaran, V., Feldman, T.~W., Anderson, W.~E., Merkle, C.~L., and Talley, D.~G., \enquote{Coupling between hydrodynamics, acoustics, and heat release in a self-excited unstable combustor,} \emph{Physics of Fluids}, Vol.~27, No.~4, 2015.
\newblock \doi{10.1063/1.4916673}.

\bibitem[{Huang et~al.(2020)Huang, Gejji, Anderson, Yoon, and Sankaran}]{Huang_2020}
Huang, C., Gejji, R., Anderson, W., Yoon, C., and Sankaran, V., \enquote{Combustion Dynamics in a Single-Element Lean Direct Injection Gas Turbine Combustor,} \emph{Combustion Science and Technology}, Vol. 192, No.~12, 2020, pp. 2371--2398.
\newblock \doi{10.1080/00102202.2019.1646732}, \urlprefix\url{https://doi.org/10.1080/00102202.2019.1646732}, doi: 10.1080/00102202.2019.1646732.

\bibitem[{Carlberg et~al.(2017)Carlberg, Barone, and Antil}]{Carlberg2017}
Carlberg, K., Barone, M., and Antil, H., \enquote{Galerkin v. least-squares {P}etrov-{G}alerkin projection in nonlinear model reduction,} \emph{Journal of Computational Physics}, Vol. 330, 2017, pp. 693--734.
\newblock \doi{10.1016/j.jcp.2016.10.033}.

\bibitem[{Arnold-Medabalimi et~al.(2022{\natexlab{a}})Arnold-Medabalimi, Huang, and Duraisamy}]{Arnold-Medalbalimi_2022}
Arnold-Medabalimi, N., Huang, C., and Duraisamy, K., \enquote{Large-eddy simulation and challenges for projection-based reduced-order modeling of a gas turbine model combustor,} \emph{International Journal of Spray and Combustion Dynamics}, Vol.~14, No. 1-2, 2022{\natexlab{a}}, pp. 153--175.
\newblock \doi{10.1177/17568277221100650}.

\bibitem[{Arnold-Medabalimi et~al.(2022{\natexlab{b}})Arnold-Medabalimi, Huang, and Duraisamy}]{Arnold-Medabalimi_2022GTMC}
Arnold-Medabalimi, N., Huang, C., and Duraisamy, K., \enquote{Large-eddy simulation and challenges for projection-based reduced-order modeling of a gas turbine model combustor,} \emph{International Journal of Spray and Combustion Dynamics}, Vol.~14, No. 1-2, 2022{\natexlab{b}}, pp. 153--175.
\newblock \doi{10.1177/17568277221100650}.

\bibitem[{Arnold-Medabalimi(2023)}]{Arnold-Medabalimi_2023}
Arnold-Medabalimi, N., \enquote{Scalable and Predictive Model Order Reduction for Reacting Flow Systems,} 2023.
\newblock \doi{10.7302/8357}, \urlprefix\url{http://deepblue.lib.umich.edu/handle/2027.42/177900}.

\bibitem[{Westbrook and Dryer(2007)}]{WestbrookDryer}
Westbrook, C.~K., and Dryer, F.~L., \enquote{Simplified Reaction Mechanisms for the Oxidation of Hydrocarbon Fuels in Flames,} \emph{Combustion Science and Technology}, Vol.~27, No. 1-2, 2007, pp. 31--43.
\newblock \doi{10.1080/00102208108946970}.

\end{thebibliography}

\end{document}